\documentclass[twocolumn]{aastex63}
\usepackage{pgfplotstable}
\usepackage{float}
\usepackage{amsmath} 

\received{Feb 1, 2021}
\revised{Mar 15, 2021}
\accepted{\today}

\usepackage{xcolor}

\submitjournal{ApJS}

\shorttitle{Background Short Period Kepler Eclipsing Binaries}
\shortauthors{Bienias et al.}

\usepackage{graphicx}
\graphicspath{{./}{figures/}}

\begin{document}

\title{Background Short Period Eclipsing Binaries in the Original \textit{Kepler} Field}

\correspondingauthor{John Bienias}
\email{john.bienias@csfk.org}

\author[0000-0002-2490-6558]{John Bienias}
\affiliation{Konkoly Observatory, Research Centre for Astronomy and Earth Sciences, E\"otv\"os Lor\'and Research Network (ELKH)\\ 
H-1121 Budapest, Konkoly Thege Mikl\'os \'ut 15-17, Hungary\\}
\affiliation{MTA CSFK Lend\"ulet Near-Field Cosmology Research Group\\}

\author[0000-0002-8585-4544]{Attila B\'odi}
\affiliation{Konkoly Observatory, Research Centre for Astronomy and Earth Sciences, E\"otv\"os Lor\'and Research Network (ELKH)\\ 
H-1121 Budapest, Konkoly Thege Mikl\'os \'ut 15-17, Hungary\\}
\affiliation{MTA CSFK Lend\"ulet Near-Field Cosmology Research Group\\}
\affiliation{ELTE E\"otv\"os Lor\'and University, Institute of Physics, Budapest, Hungary\\}

\author[0000-0001-9394-3531]{Adrienn Forr\'o}
\affiliation{Konkoly Observatory, Research Centre for Astronomy and Earth Sciences, E\"otv\"os Lor\'and Research Network (ELKH)\\ 
H-1121 Budapest, Konkoly Thege Mikl\'os \'ut 15-17, Hungary\\}
\affiliation{E\"otv\"os Lor\'and University, P\'azm\'any P\'eter s\'et\'any 1/A, Budapest, Hungary}
\affiliation{MTA CSFK Lend\"ulet Near-Field Cosmology Research Group\\}

\author[0000-0001-8060-2367]{Tam\'as Hajdu}
\affiliation{Konkoly Observatory, Research Centre for Astronomy and Earth Sciences, E\"otv\"os Lor\'and Research Network (ELKH)\\ 
H-1121 Budapest, Konkoly Thege Mikl\'os \'ut 15-17, Hungary\\}
\affiliation{E\"otv\"os Lor\'and University, P\'azm\'any P\'eter s\'et\'any 1/A, Budapest, Hungary}
\affiliation{MTA CSFK Lend\"ulet Near-Field Cosmology Research Group\\}

\author[0000-0002-3258-1909]{R\'obert Szab\'o}
\affiliation{Konkoly Observatory, Research Centre for Astronomy and Earth Sciences, E\"otv\"os Lor\'and Research Network (ELKH)\\ 
H-1121 Budapest, Konkoly Thege Mikl\'os \'ut 15-17, Hungary\\}
\affiliation{MTA CSFK Lend\"ulet Near-Field Cosmology Research Group\\}
\affiliation{ELTE E\"otv\"os Lor\'and University, Institute of Physics, Budapest, Hungary\\}

\begin{abstract}
During the primary Kepler mission, between 2009 and 2013, about 150,000 pre-selected targets were observed with a 29.42 minute-long cadence. However, a survey of background stars that fall within the field of view (FOV) of the downloaded apertures of the primary targets has revealed a number of interesting objects. In this paper we present the results of  this search focusing on short period eclipsing binary (SPEB) stars in the background pixels of primary Kepler targets. We used Lomb-Scargle and Phase Dispersion Minimisation methods to reveal pixels that show significant periodicities, resulting in the identification of 547 previously unknown faint SPEBs, mostly W UMa type stars, and almost doubling the number of SPEBs in the original Kepler FOV. We prepared the light curves for scientific analysis and cross matched the pixel coordinates with Gaia and other catalogues to identify the possible sources. We have found that the mean of the brightness distribution of the new background SPEBs is $\sim4-5$ magnitudes fainter than other, primary target eclipsing binaries in the Kepler Eclipsing Binary catalogue. The period distribution nonetheless follows the same trend but the spatial distribution appears to be different from that described by \citet{kirk2016} for the catalogue eclipsing binaries.
\end{abstract}


\section{Introduction} \label{sec:intro}

Eclipsing binary systems provide a unique opportunity for obtaining fundamental stellar parameters including mass, radius, luminosity and temperature. Because the components' ages and chemical compositions are the same, investigation of eclipsing binaries is also essential for testing stellar evolutionary theories. In recent decades, surveys such \textit{OGLE} \citep{OGLE1997}, \textit{CoRoT} \citep{CoRoT2009}, \textit{Kepler} \citep{borucki2010} and the ongoing \textit{TESS} \citep{TESS2015} have contributed to a dramatic rise in the number of known eclipsing binaries. Of these, the \textit{Kepler} telescope in particular provides a major step forward in the detailed study of binary systems due to its ultra-high precision photometry, including highlights such as Kepler-47 where three planets have been detected orbiting a binary star \citep{Orosz2019}.

The \textit{Kepler} photometric space telescope was launched in 2009 into an Earth-trailing heliocentric orbit and was designed to detect exoplanets by observing around 150,000 target stars in a fixed 105 square degree area of the sky in the Cygnus, Lyra and Draco constellations. The observations were made with a 29.42 minute (long) cadence over a period of 17 Quarters from 2009 to 2013 at which time a second reaction control wheel failed and this part of the mission was terminated. For telemetry bandwidth reasons, only pixels of the target stars and their immediate surroundings were downloaded, and analysis of these images has focused on the pixels within the optimal apertures of the target stars. However, a pixel-by-pixel analysis of these images reveals a variety of interesting objects in the background.

The latest \textit{Kepler} Eclipsing Binary Catalogue \footnote{Obtainable from \url{http://keplerebs.villanova.edu/}} \citep{kirk2016} contains 2922 eclipsing binaries identified amongst the \textit{Kepler} main targets. By searching the \textit{Kepler} background stars, it was hoped to extend this number considerably, and this paper presents the results of a search of the \textit{Kepler} observations restricted to short period eclipsing binaries (SPEBs), that is, with an orbital period up to 12 hours. 547 such objects have been found, mostly W UMa types, but also a number of Algol, $\beta$ Lyrae and ellipsoidal types.

Note that reference is made throughout this paper to the main target eclipsing binaries listed in the \textit{Kepler} Eclipsing Binary Catalogue and to the new (non-target) eclipsing binaries described in this work. For the avoidance of confusion, where it is necessary to distinguish between them, the main target catalogue eclipsing binaries will be referred to as cat-EBs, and the non-target SPEBs as nt-EBs. In addition, KIC main target numbers are appended with ''nt-EB'' where appropriate to make it clear that reference is being made to a SPEB discovered in the KIC aperture, not the KIC target itself.

Short period eclipsing binaries are important for investigating the evolution of low-mass stars and in particular for investigation of the well-known orbital period cutoff at approximately 0.22 days, below which few binaries are observed. \citet{Stepien2006} attributes this to the time required for magnetic wind-driven angular momentum losses to evolve a detached binary into a contact binary. The time required to obtain a contact binary with period less than the cutoff is greater than the age of the universe.

In this paper we report the results of a search for background short period eclipsing binaries in the immediate vicinity of all the primary Kepler targets in the original mission. In Sec.~\ref{sec:processing} we discuss data processing, including period search methods, the identification and classification of our candidates, while in Sec.~\ref{sec:results1} we discuss their physical parameters, spatial distribution, and hints of eclipse timing variation in the sample. In Sec.~\ref{sec:summary} we give a brief summary and in Appendices~\ref{sec:results2} and \ref{sec:phys_param} we present the properties of the newly found short period eclipsing binary systems.

Findings for longer period eclipsing binaries and RR Lyrae type variable stars will be published in separate papers.

\section{Processing of \textit{Kepler} Data} \label{sec:processing}
\subsection{Identification of SPEB candidates}
The \textit{Kepler} observations were made quasi-continuously over a 4 year period. However, the data is provided in discrete Quarterly sets due to the 90-degree rolls required to maintain the correct orientation of the solar panels.
Our initial search was restricted to the Quarter 4 (Q4) observations, since this was the first relatively quiet Quarter. All the individual pixels belonging to long cadence Q4 KIC target apertures were used.
The light curves were extracted for each individual pixel for each main target (denoted by Kepler Input Catalog numbers) and a low-resolution Lomb-Scargle (LS; \citealt{Lomb76} and \citealt{Scargle82}) algorithm with a frequency interval $\approx 0.01$ cycles/day employed to generate a periodogram for each pixel light curve. A more detailed description of the method is described in an accompanying paper (Forr\'o et al. in preparation).

\begin{figure}
\centering
    \includegraphics[width=\columnwidth]{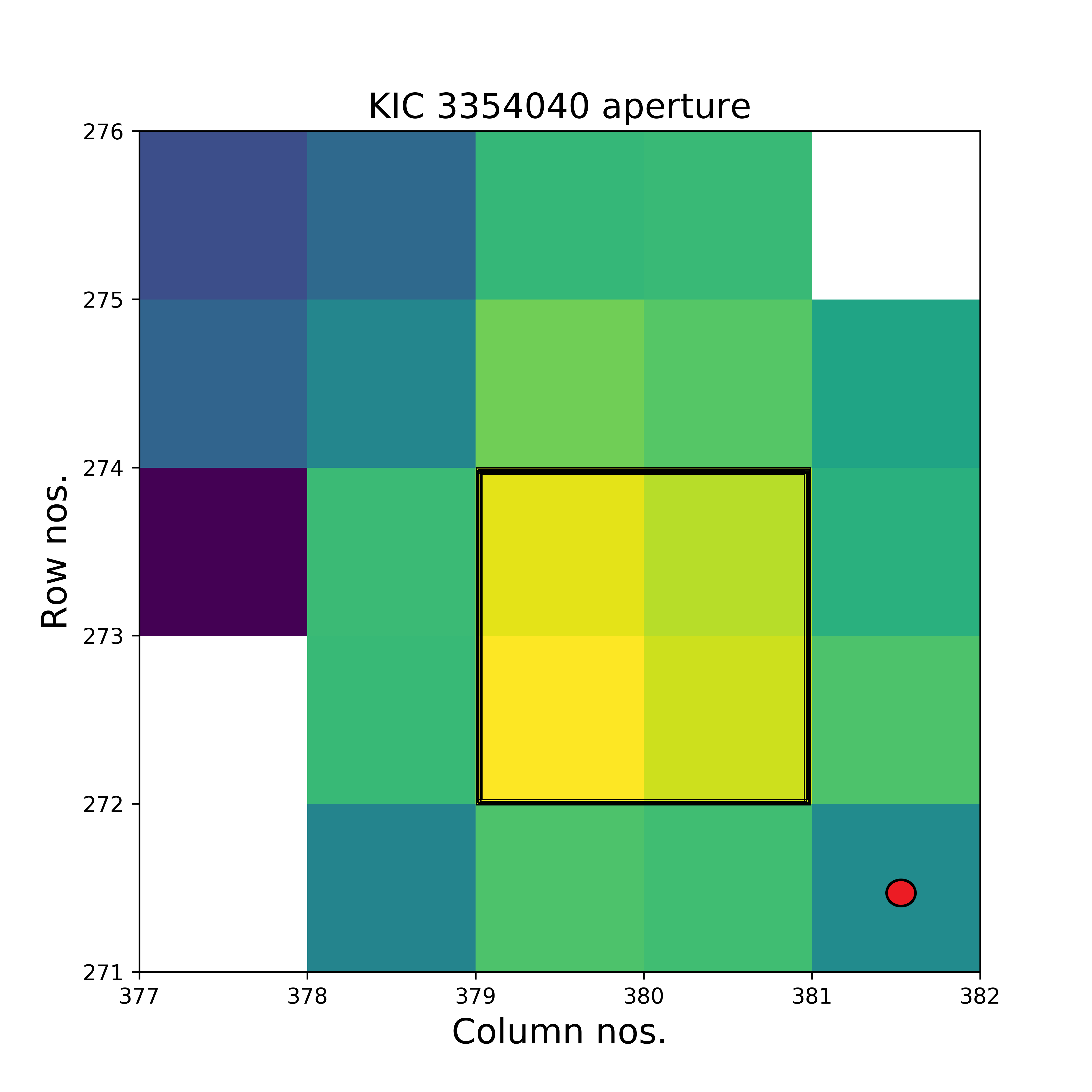}
    \includegraphics[width=\columnwidth]{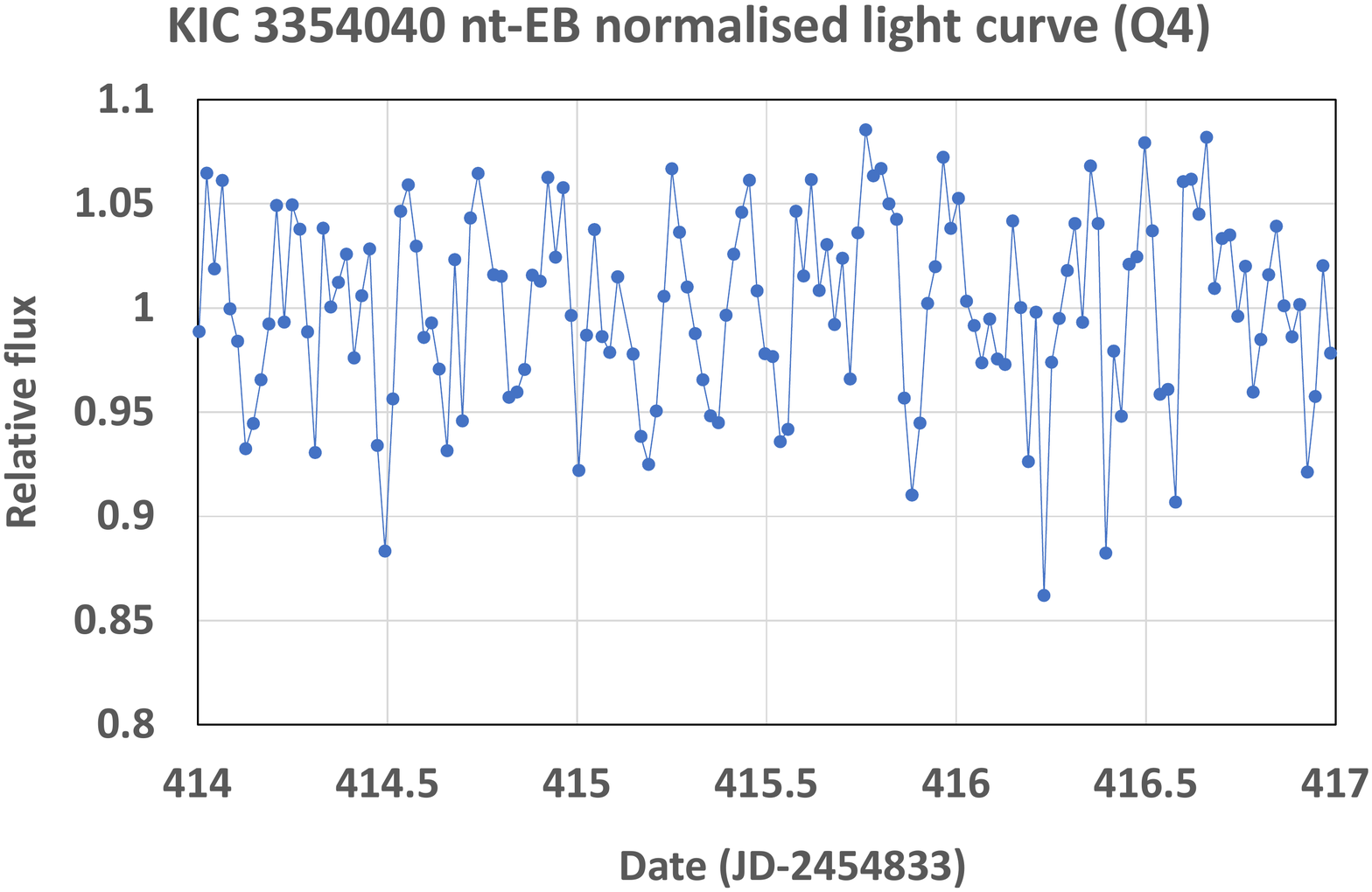}
    \includegraphics[width=\columnwidth]{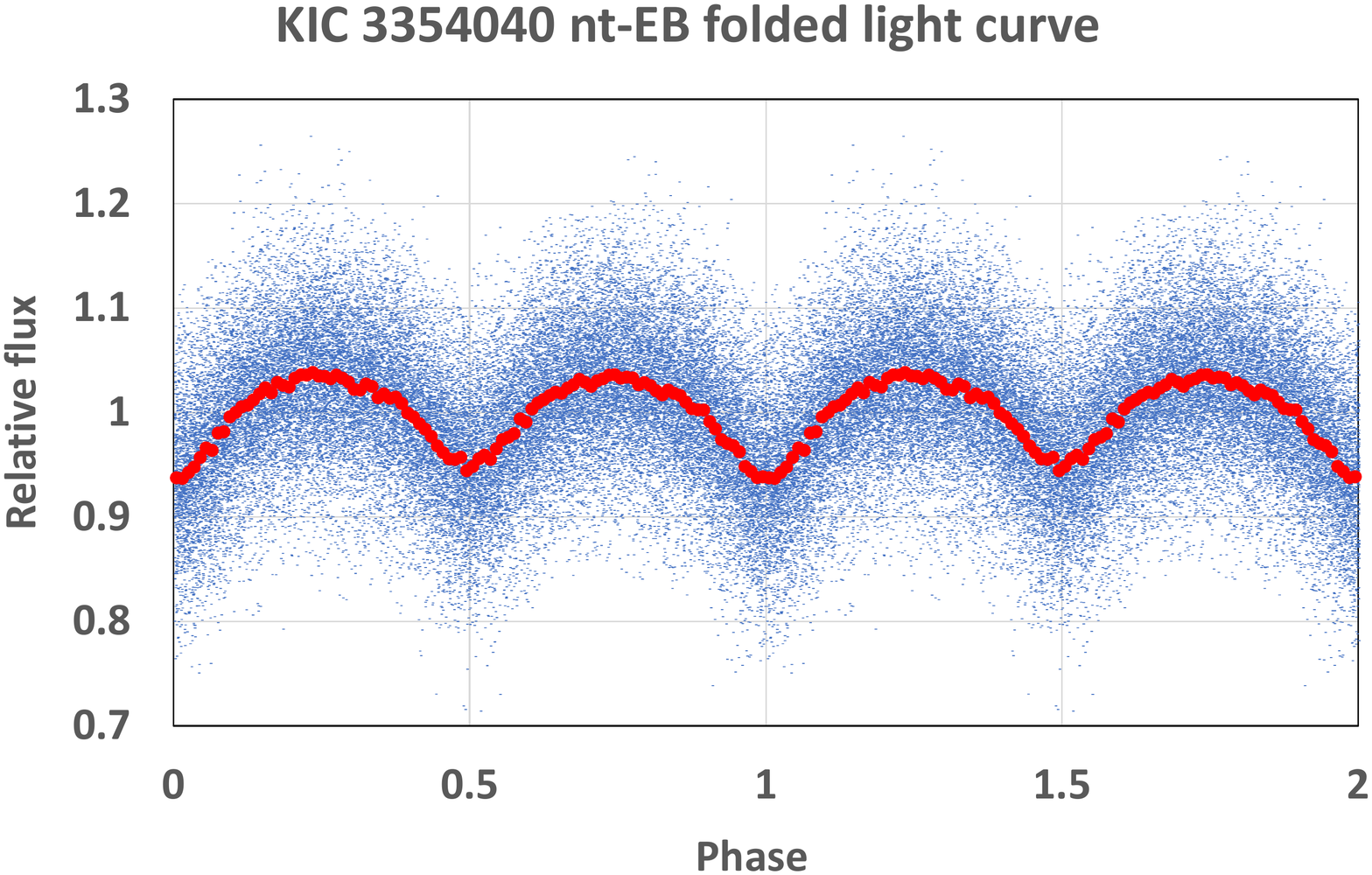}
\caption{KIC 3354040 aperture mask (\textit{top}) with the nt-EB pixel marked with a red circle. (source: Lightkurve). The colours indicate the flux level, with blue being a low flux and yellow a high flux. The four bright pixels used in the \textit{Kepler} pipeline to extract the main target light curve are outlined in black. The partial normalised light curve for the nt-EB pixel is shown in the  \textit{middle} and the combined Q2 - Q16 folded light curve is at the \textit{bottom}. This shows the individual observations (\textit{blue}) and the phase bin average fluxes (\textit{red}).}
\label{fig:3354040}
\end{figure}

The resulting periodograms were searched for individual pixels showing a significant  dominant frequency greater than 4 cycles/day. 'Significant' means the amplitude is greater than 4 times the local background, that is: 
\begin{equation}
a_{d} - a_{ave} \geq 4 \times d_{rms}
\end{equation}
where $a_{d}$ is the dominant frequency amplitude, $a_{ave}$ is the average amplitude of the frequencies within $\pm 0.5$ c/d of the dominant frequency, and $d_{rms}$ is the RMS deviation from $a_{ave}$ of the frequencies within $\pm 0.5$ c/d of the dominant frequency.

After elimination of known objects, false positives (arising from bleeding, ghosting or crosstalk), pulsating and rotational variables and duplicates (where a candidate PSF is detected in two KIC apertures), the individual pixel light curves for each remaining candidate were merged and detrended. The detrending procedure is described briefly in section \ref{sub:detrend_proc} below. On average, six pixels were found for each candidate, though the number varies widely. Many of the Q4 pixels found show only a weak and noisy signal either because the source binary is faint (the source candidates have an average Gaia G magnitude $\approx 18.2$) or lies far from the KIC aperture so that only the edge of the PSF is captured. An example is given in Fig.~\ref{fig:3354040}, showing the single nt-EB pixel found on the KIC~3354040 Q4 aperture. The four bright pixels used in the \textit{Kepler} pipeline to extract the main target light curve are outlined in black. The top right pixel is blank because no data was downloaded from that pixel and the two pixels at the bottom left have been left blank because the flux is given on a logarithmic scale and these two pixels have negative flux values. A portion of the associated normalised light curve and the combined Q2 - Q16 phase-folded light curve for the nt-EB are also shown. The candidate source for this binary is Gaia EDR3 2052119499831397248 which has a Gaia G magnitude of 18.451 and lies approximately $5\arcsec$ from the location of the marked pixel. The source identification process is described in section \ref{sub:source_id}.

For each candidate, a search was carried out on the relevant Q2 - Q16 pixel light curves to find the pixels with the same dominant frequency (to within $\pm0.01$ cycles/day) as the identified candidates. Quarters 0, 1 and 17 were discarded as they are truncated and give rise to a disproportionate number of frequencies that differ significantly from those obtained from Quarters 2 - 16. The resulting pixel light curves were then merged, detrended and normalised to provide a set of Quarterly light curves in flux.

\subsection{Detrending Procedure} \label{sub:detrend_proc}

The Quarterly light curves are subject to discontinuities, and so each light curve was split into segments which could be detrended separately. Each such segment was then detrended and normalised by fitting, and dividing by, a fourth order polynomial and the individual segments were then re-combined into a single light curve. Each resulting light curve was examined for spikes and other poorly fitting observations and these were removed.

The Quarterly light curves thus obtained were combined into a single Q2 - Q16 light curve with each Quarter scaled to the Quarter showing the highest level of normalised flux.

\subsection{Determination of Orbital Periods} \label{sub:orb_period}

\begin{figure}
    \centering
    \includegraphics[width=\columnwidth]{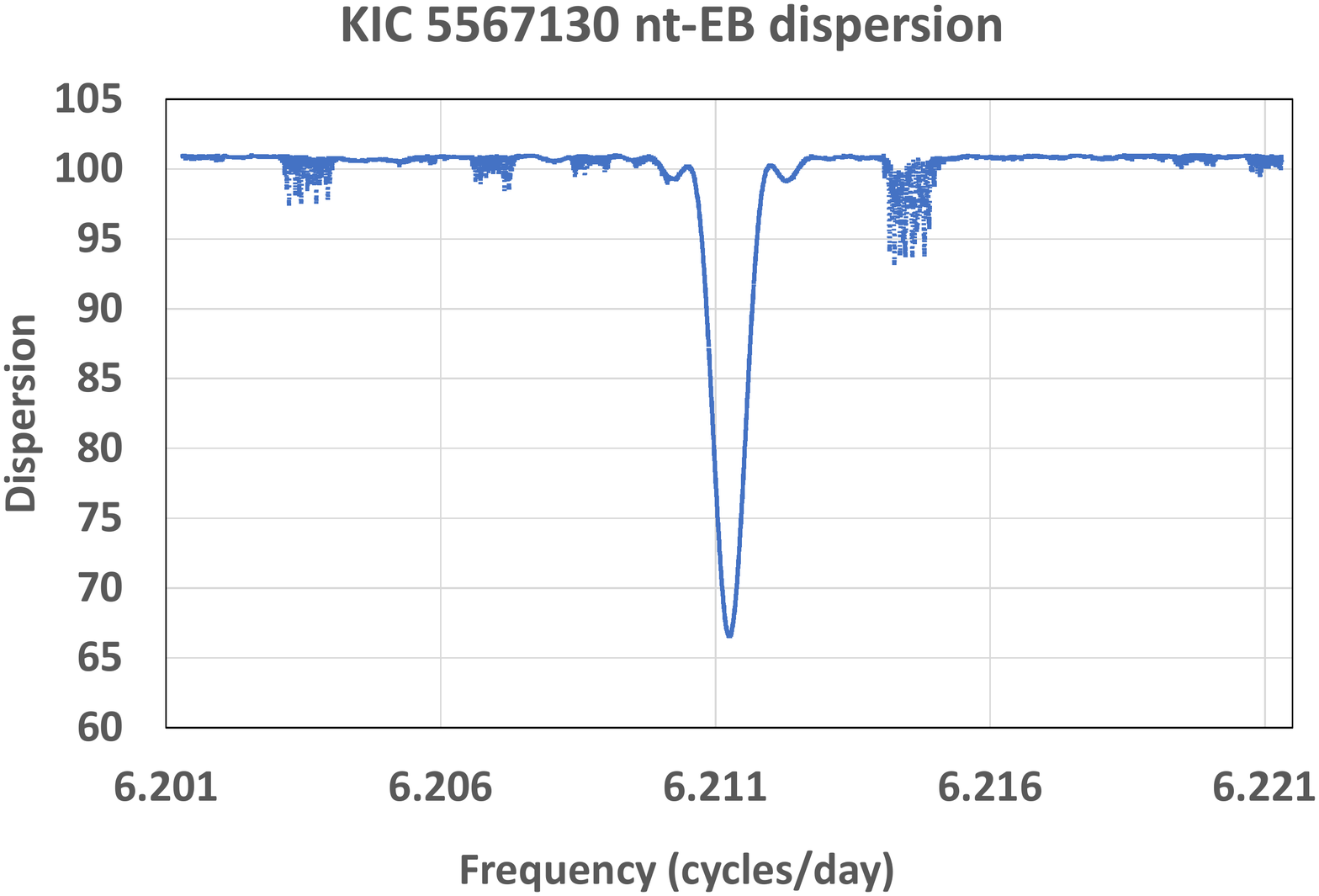}
    \includegraphics[width=\columnwidth]{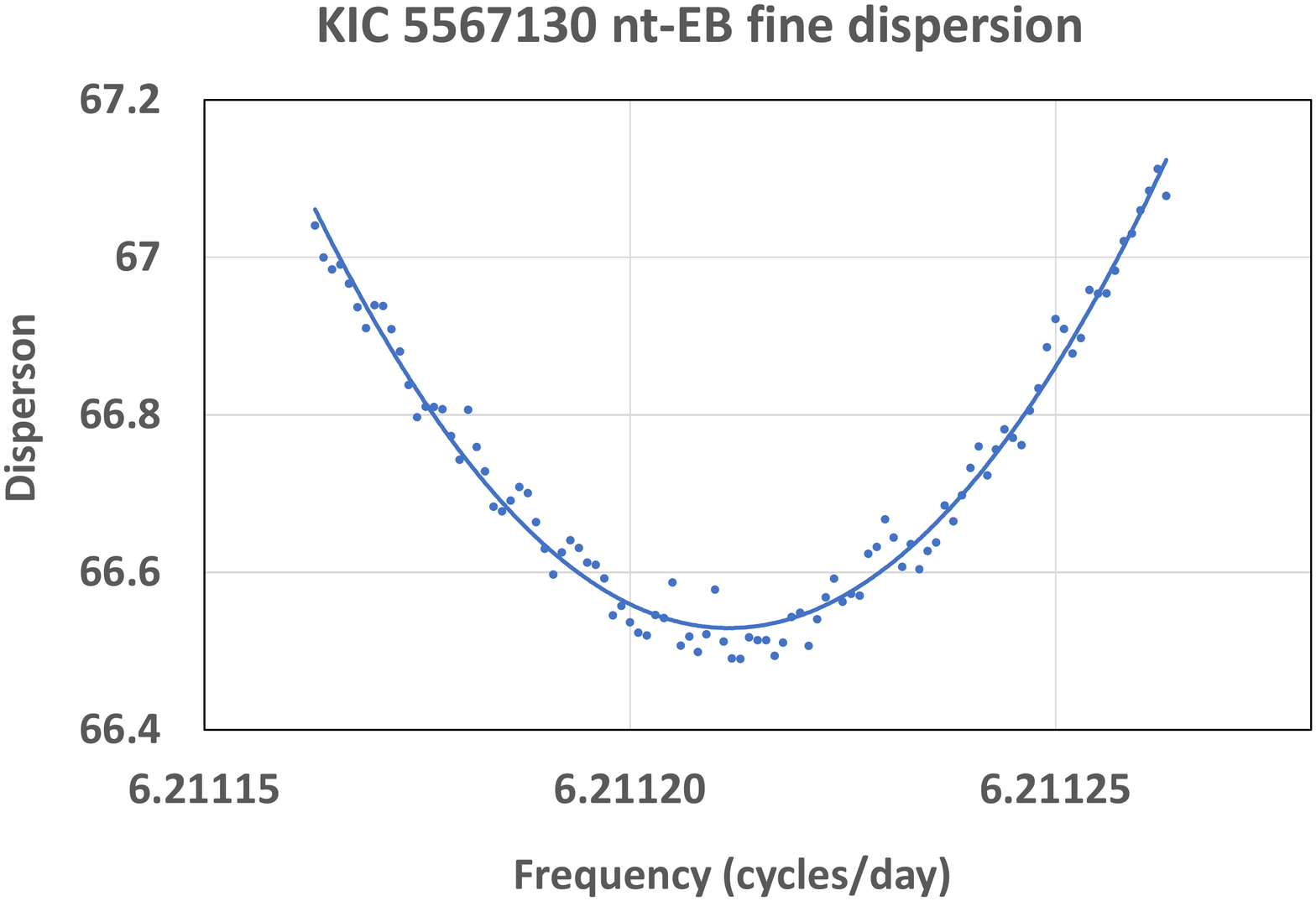}
    \includegraphics[width=\columnwidth]{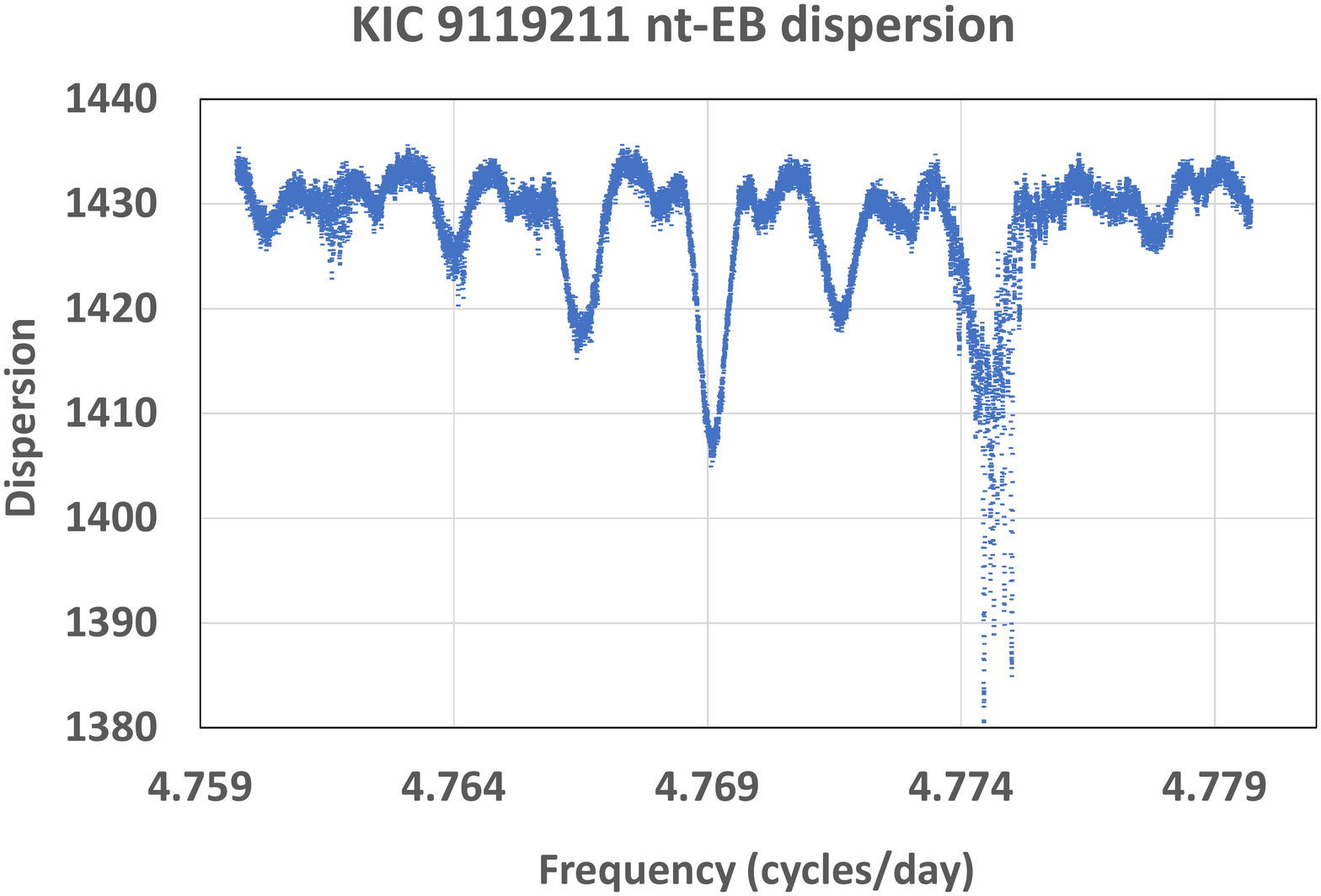}
    \caption{The \textit{top} figure shows the variation of PDM dispersion with frequency for a good quality light curve. The \textit{middle} figure shows the area around the minimum, fitted with a parabola and the \textit{bottom} figure shows the dispersion curve for a poor quality light curve, with a false minimum at frequency $\approx 4.775$ cycles/day. The true frequency $\approx 4.769$ cycles/day.}
    \label{fig:dispersion}
\end{figure}

As the Lomb-Scargle algorithm initially used is low-resolution and does not provide precise frequencies, the frequencies obtained were progressively improved, as follows:
\begin{enumerate}
    \item For each Quarterly light curve, a Phase Dispersion Minimization (PDM) algorithm was employed to search for the frequency (in a range of \(\pm0.01\) cycles/day around the LS frequency) which yields the minimum dispersion (scatter) for the phase-folded light curve, that is, the minimum value of:
    
\begin{equation}
\sum_{i=1}^{n}\sum_{j=1}^m(O_{ij}-E_i)^{2}/E_i
\end{equation}

where $n$=30 is the number of phase bins, $O_{ij}$ is the ${j}^{th}$ observed flux value for the ${i}^{th}$ bin and $E_i$ is the average flux value for the ${i}^{th}$ bin. In many cases, the Quarterly light curves were of poor quality and the frequencies obtained unreliable. Thus, frequencies more than two standard deviations from the average Quarterly frequency for the candidate were discarded and the average recalculated. The average frequency thus obtained was used as the starting point for the following step. Fig.~\ref{fig:6105113} shows the Quarterly LS and PDM frequencies for the nt-EB in the KIC 6105113 aperture (candidate Gaia EDR3 2104040161180070784), illustrating the typical level of frequency improvement. The standard deviation of the PDM frequencies is, on average, approximately 18 times smaller than that of the LS frequencies.

\begin{figure}
  \centering
  \includegraphics[width=\columnwidth]{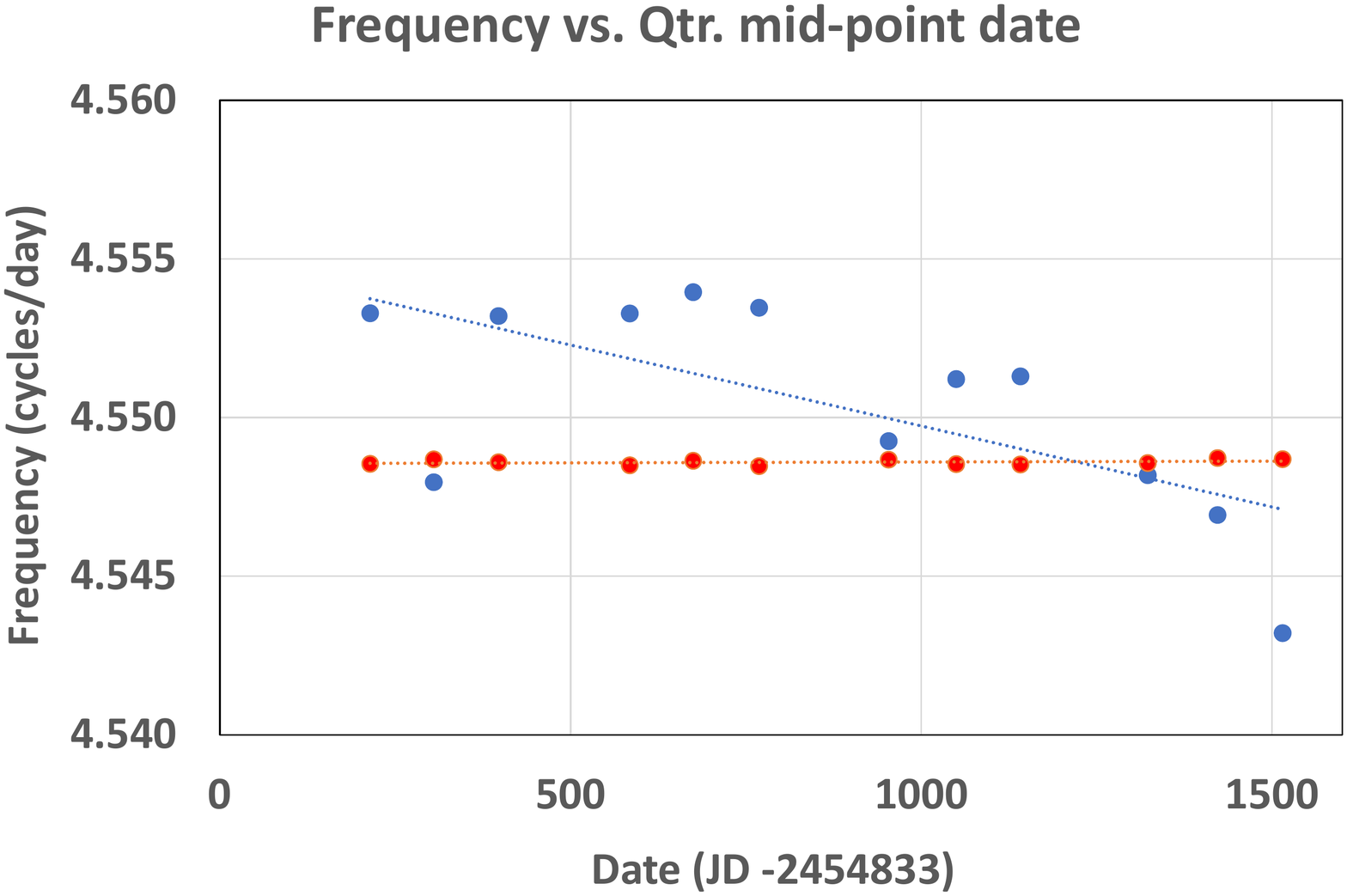}
  \caption{Quarterly LS frequencies (\textit{blue}) and PDM frequencies (\textit{red}) obtained for an nt-EB in the KIC 6105113 aperture.}
  \label{fig:6105113}
\end{figure}

   \item The same PDM algorithm was employed, but with 500 phase bins and using the Q2 - Q16 combined light curve to search for the best frequency in a range of \(\pm0.01\) cycles/day around the starting frequency. A fine adjustment was then made to the resulting frequency by fitting a parabola to the dispersion curve close to the minimum value. In six cases where the light curve is of poor quality, the PDM method generated false minima. Fig.~\ref{fig:dispersion} shows examples of dispersion variation by frequency. The \textit{top} chart shows a typical dispersion curve, with a clearly defined minimum. The fit to the fine dispersion curve is shown in the \textit{middle} chart and the \textit{bottom} chart shows the dispersion curve for a poor quality light curve. This has a false minimum at frequency $\approx 4.775$ cycles/day, while the true minimum is at frequency $\approx 4.769$ cycles/day. In these six cases, the correct frequency was obtained manually from the dispersion data and then the fine adjustment applied.
   \item A standard high-resolution Lomb-Scargle algorithm was also employed to find the best frequency for the combined Q2 - Q16 light curves in the $\pm 0.1$ cycles/day range around the PDM frequency obtained in the previous step.
\end{enumerate}
Typically, the PDM method provides better results for Algol type binaries while the LS methodology provides better results for sinusoidal light curves, i.e. W UMa and ellipsoidal light curves. 

Both sets of orbital periods obtained are shown in Table \ref{tab:results}. These are in good agreement, with all except one agreeing to within $10^{-5}$ days and 94.5\% agreeing to within $10^{-6}$ days. The one exception is the Algol type nt-EB in the KIC 9520443 aperture, where the difference is $1.56\times10^{-5}$ days.

It should be noted that five Algol type nt-EBs were found with true period equal to twice the initial LS period. Two of these have periods $\sim0.51$ days and the other three have periods $\sim0.9$ days. Therefore, in comparisons with the \textit{Kepler} EB catalogue, the effective range of the nt-EBs is deemed to be $0.18 - 0.52$ days, and the equivalent set of cat-EBs is those within the same orbital period range.

\subsection{Determination of ephemerides  and eclipse depths} \label{sub:ephemeris}
The phase-folded and binned light curves were used to determine the ephemeris times of the nt-EB candidates. The two continuous areas with flux below the normalised value of 1 were selected as potential eclipses and an eighth-order polynomial was fitted to each such eclipse to determine their phase and flux values, using the Newton-Raphson method. The lowest point of the deeper eclipse was assigned a phase value of zero, and the last date before the start of the light curve corresponding with this phase was assigned to the zero epoch, $BJD_0$. These values are shown in Table \ref{tab:results}.

\begin{figure}
    \centering
    \includegraphics[width=\columnwidth]{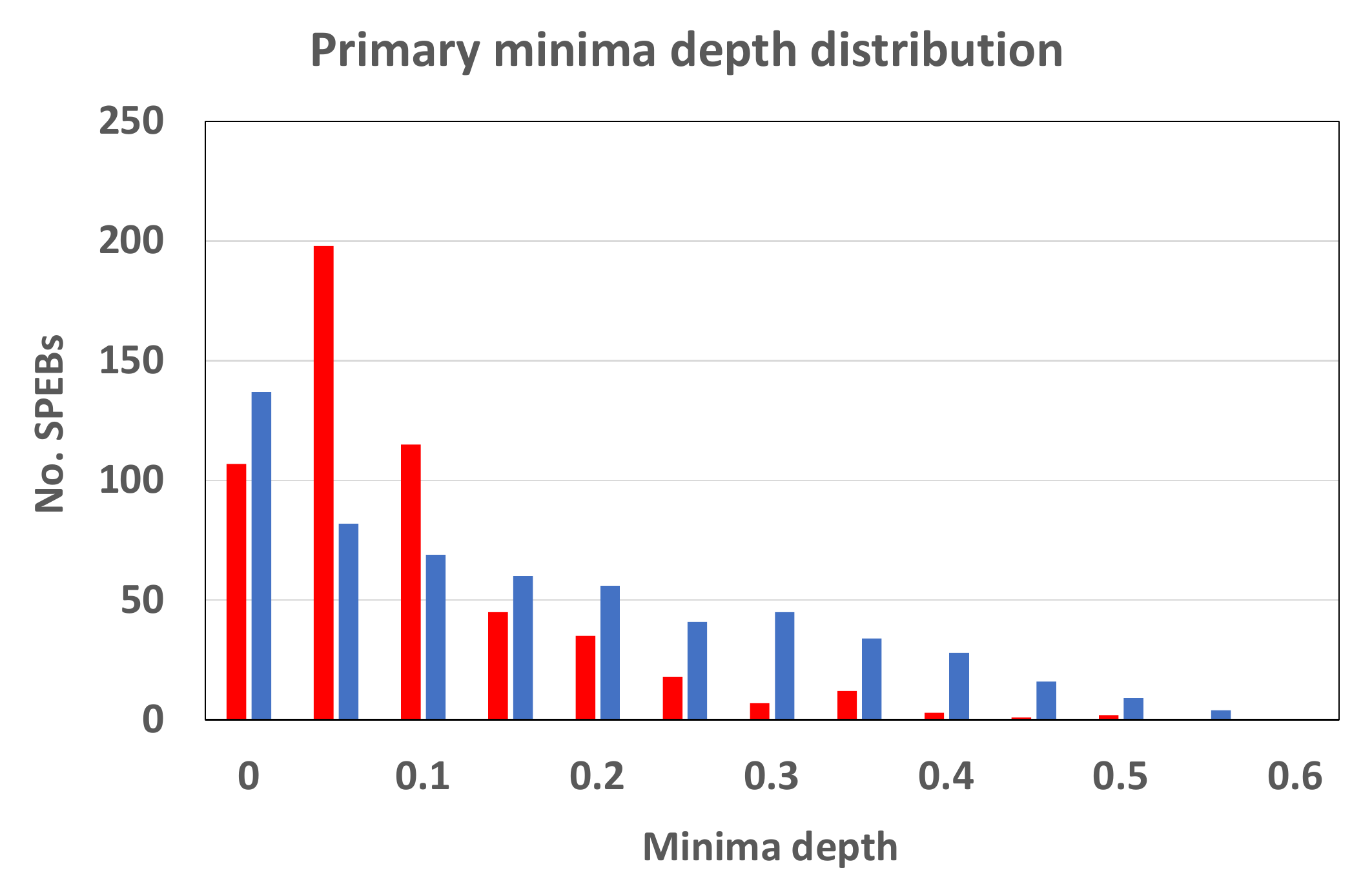}
    \includegraphics[width=\columnwidth]{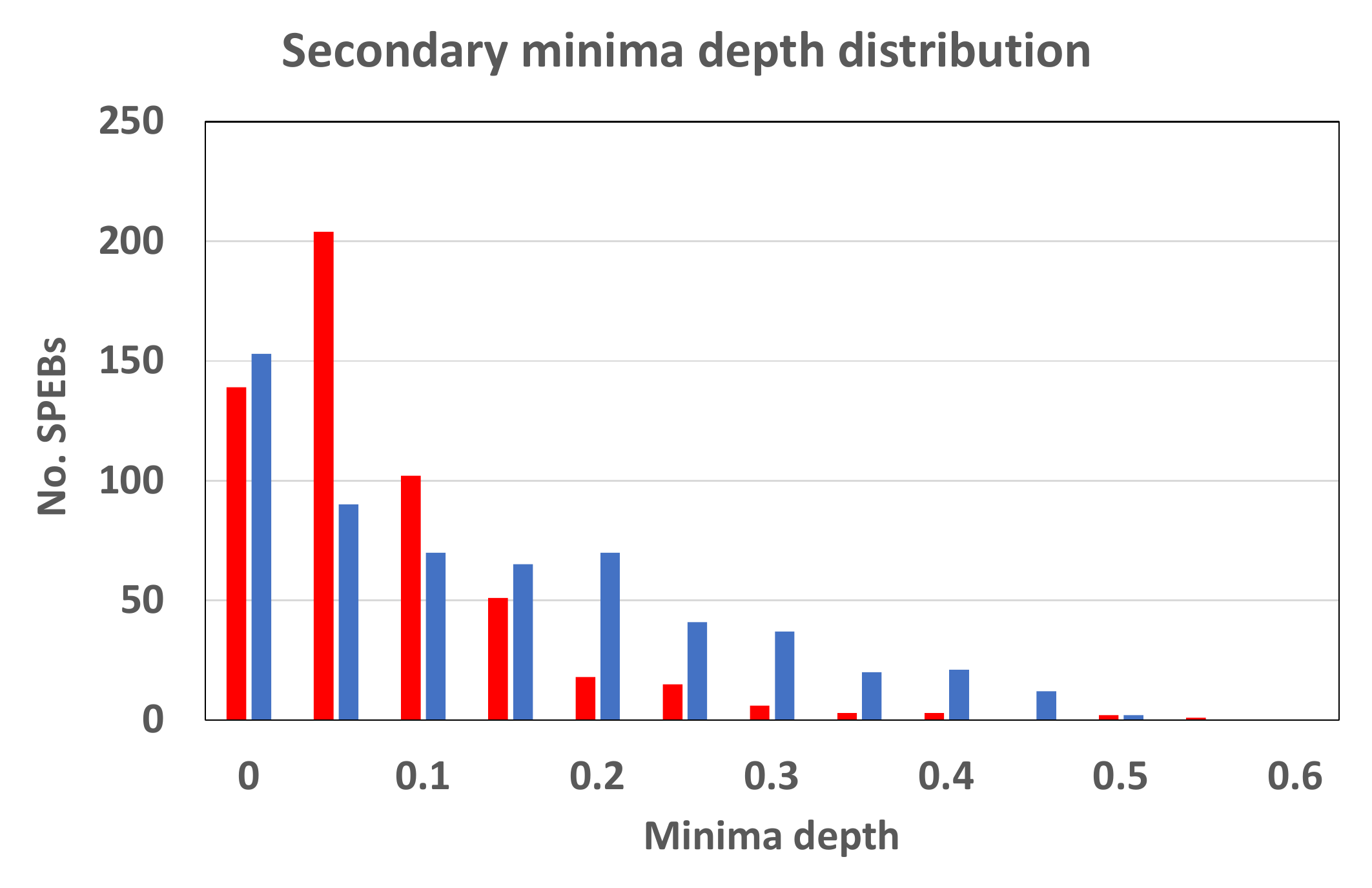}
    \includegraphics[width=\columnwidth]{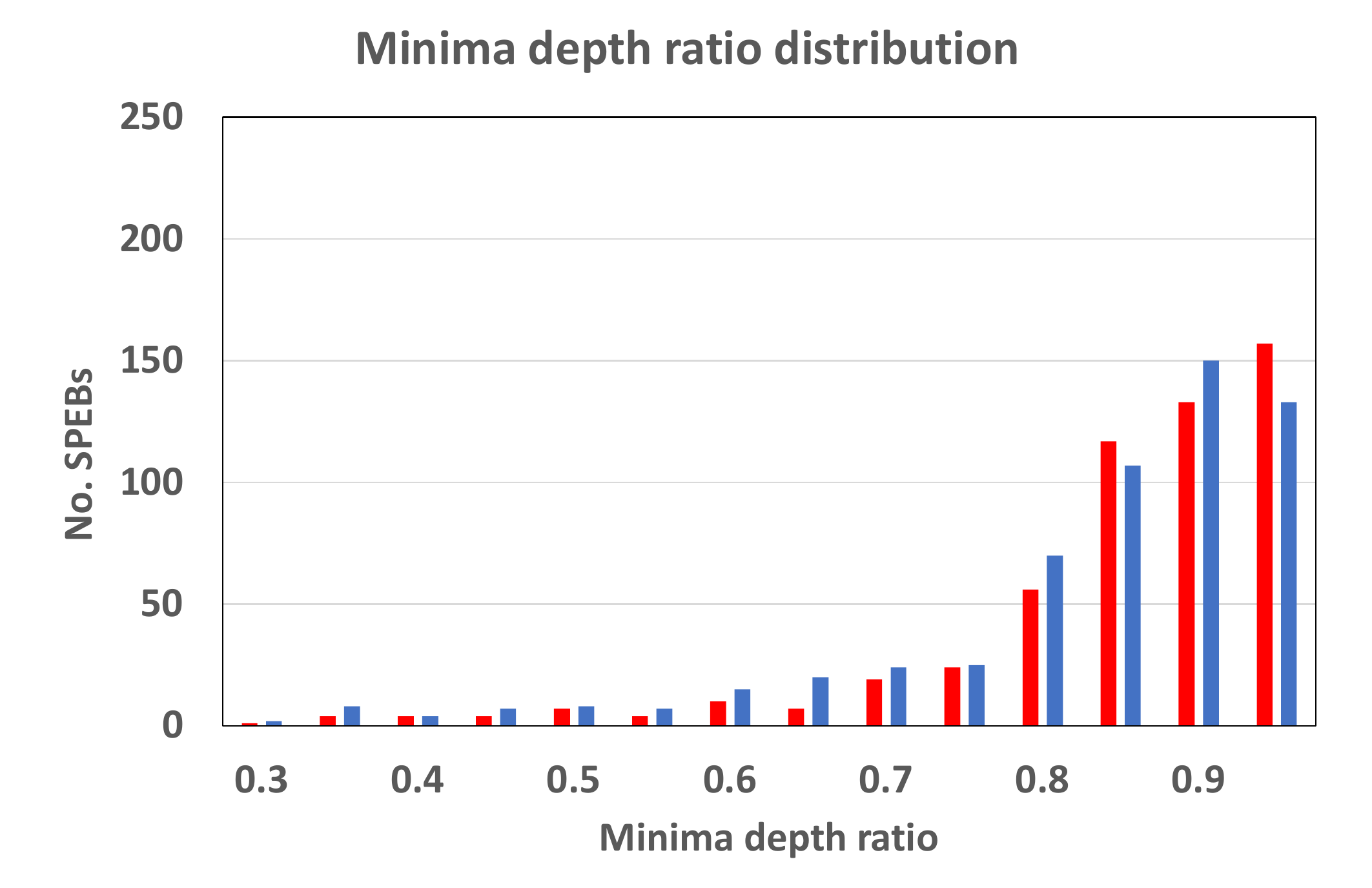}    
    \caption{The \textit{top} figure shows the distribution of primary minima depths for the nt-EBs (\textit{red}) and for the equivalent set of cat-EBs (i.e. with $0.18 < P < 0.52$ days) (\textit{blue}). The \textit{middle} figure shows the distribution of the secondary minima depths and the \textit{bottom} shows the distribution of the secondary/primary depth ratios. The minima depth and minima ratio axis labels show the minimum value for each bin.}
    \label{fig:eclipse_depth}
\end{figure}

The light curve maxima were calculated with a similar method and the eclipse depths determined from the differences between the maxima and the minima. The eclipse depth distributions of the systems are presented in Fig. \ref{fig:eclipse_depth}. The same calculations were performed on the equivalent set of EBs from the \textit{Kepler} EB catalogue, i.e. with $0.18 < P < 0.52$ days, and these are also shown in Fig.~\ref{fig:eclipse_depth} for comparison.

It will be seen that the primary and secondary eclipse minima (in the \textit{top} and \textit{middle} charts respectively) for the nt-EBs are skewed towards shallower depths relative to the cat-EBs. This is attributed to the relative faintness of the nt-EBs and their proximity to brighter stars such that the contamination from these nearby stars has the effect of reducing the apparent eclipse depths. This is supported by the \textit{bottom} chart showing the eclipse depth ratio distributions which are similar for the nt-EBs and cat-EBs. Note that the minima and maxima depths are normalised, i.e. a complete loss of light would have a depth of 1.0.

However, this minima skew does not hold for the lowest $0-0.05$ minima depth bin, where it is assumed that the search for SPEBs has failed to find a proportion of very small eclipses among the fainter background stars.

\subsection{Classification of candidates} \label{sub:class}

The phase-folded light curves were initially categorised by visual inspection into Algol (EA), $\beta$ Lyrae (EB), W UMa (EW) or Ellipsoidal (ELL) types. Additionally, a systematic classification was carried out based on the light curve morphology. Following the solution of \citet{KeplerLLE}, a Local Linear Embedding (LLE; \citealt{LLEalgo}) algorithm was used to cluster the four second-order polynomial chain (polyfit;  \citealt{polyfit}) representation of the light curves. First, the polyfits were projected into a 3 dimensional space using LLE, then this manifold was projected into a 2D space using another LLE. The resultant two axes form an arc, which can be represented as a one dimensional curve. This was transformed into the range of 0--1, which is referred to as the morphology parameter \textit{c}, which were used to separate the sub-classes. A low value of this parameter, close to zero, indicates an Algol type and a high value, close to one, indicates a W UMa or ellipsoidal type. The details of this process are published in an analysis of OGLE IV data by \citet{BodiLLE}.

Table \ref{tab:class} shows the range of the morphological classification parameter, $c$, for each visual classification. The ranges overlap somewhat, indicating the difficulty of assigning classifications on the basis of photometric light curves alone, particularly for the ELL and EW types. 

\begin{table}
\begin{tabular}{ |c|c| }
\hline
 Visual Class & Morphological Class, $c$ \\
 \hline
 ELL& 0.751 -- 0.988 \\ 
 EW & 0.622 -- 0.983 \\ 
 EB & 0.566 -- 0.702 \\ 
 EA & 0.441 -- 0.587 \\
 \hline
\end{tabular}
\caption{Comparison of visual classification and morphological classification parameter, $c$.}
\label{tab:class}
\end{table}

\begin{figure}
  \centering
  \includegraphics[width=\columnwidth]{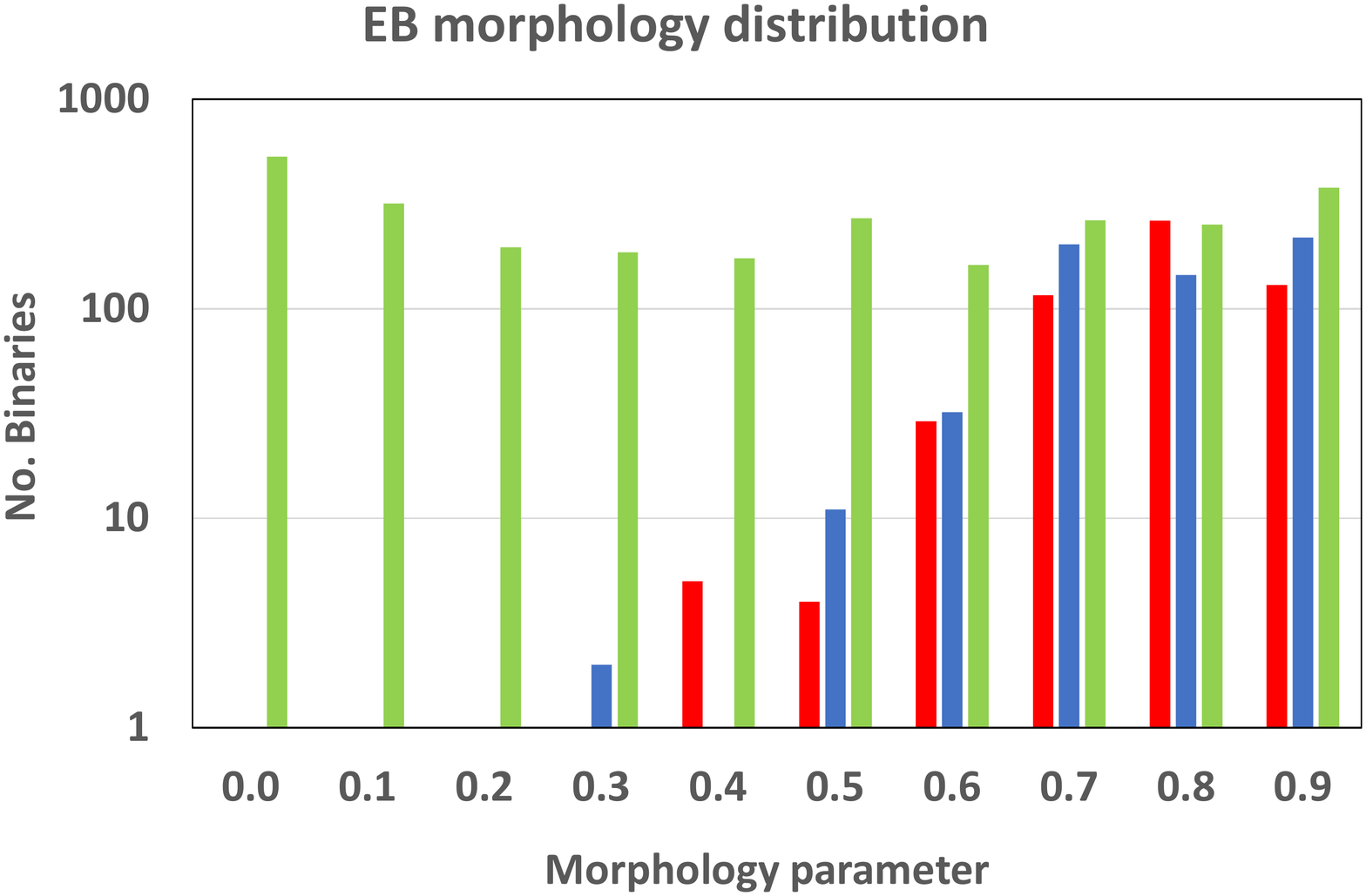}
  \caption{Comparison of \textit{Kepler} EB catalogue morphologies with the nt-EBs presented in this work. The figure shows the distribution of nt-EBs (\textit{red}), the distribution of the equivalent period cat-EBs (\textit{blue}) and the distribution of the catalogue EBs of all periods (\textit{green}). The morphology parameter axis labels denote the minimum value for each bin.}
  \label{fig:EB_morph}
\end{figure}

The parameter, $c$, is similar to the morphology parameter associated with the \textit{Kepler} EB catalogue entries, and a comparison between the two is given in Fig.~\ref{fig:EB_morph}, showing the distribution of the various EBs by morphology parameter. The nt-EBs (\textit{red}) and the equivalent set of cat-EBs (\textit{blue}) have broadly similar distributions concentrated in the $0.6 - 1.0$ range (W UMa and ellipsoidal types), as might be expected with short period, and thus close binaries. The distribution for the entire set of cat-EBs is also shown (\textit{green}).

\begin{figure}
\centering
    \includegraphics[width=\columnwidth]{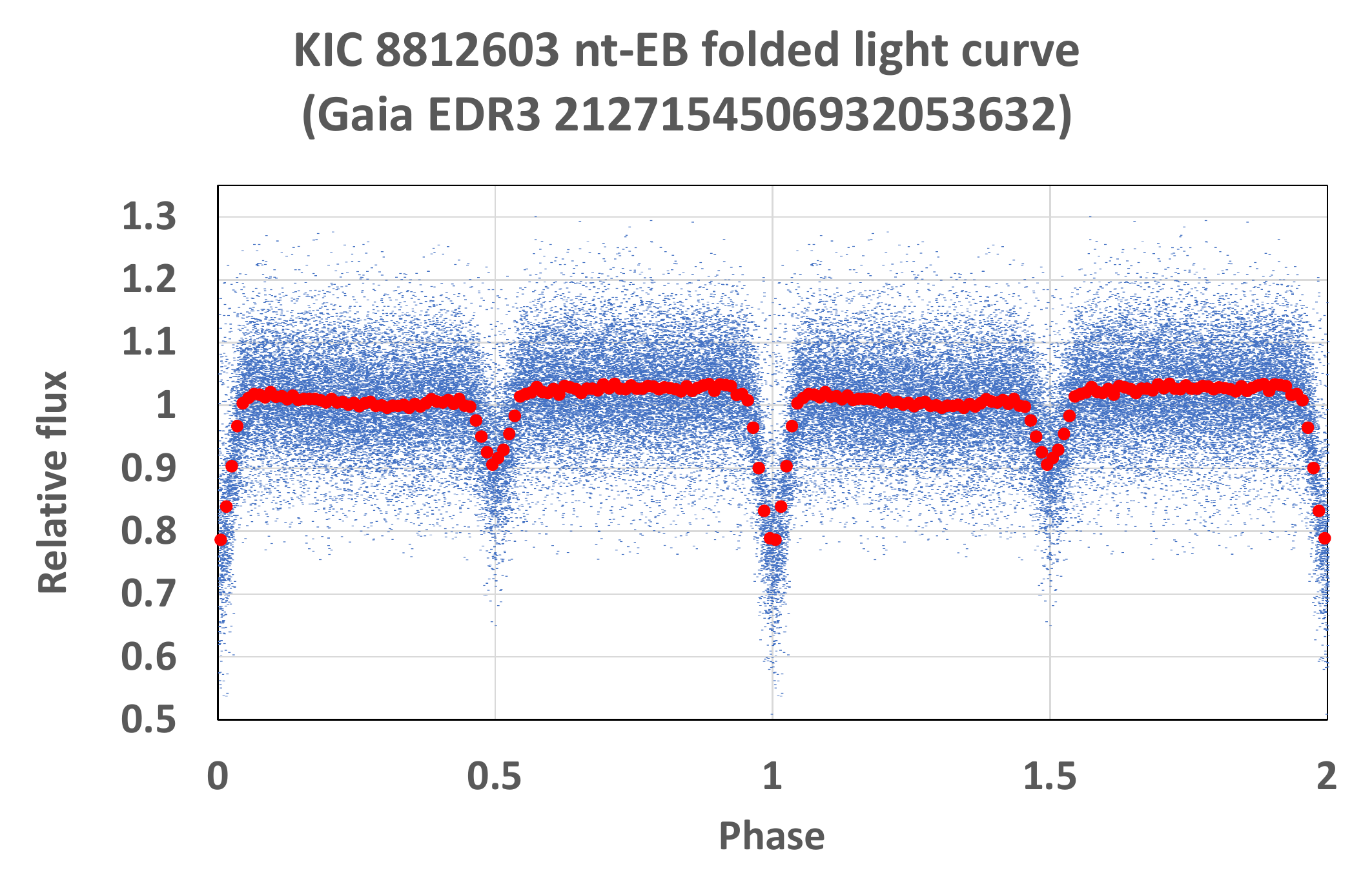}
    \includegraphics[width=\columnwidth]{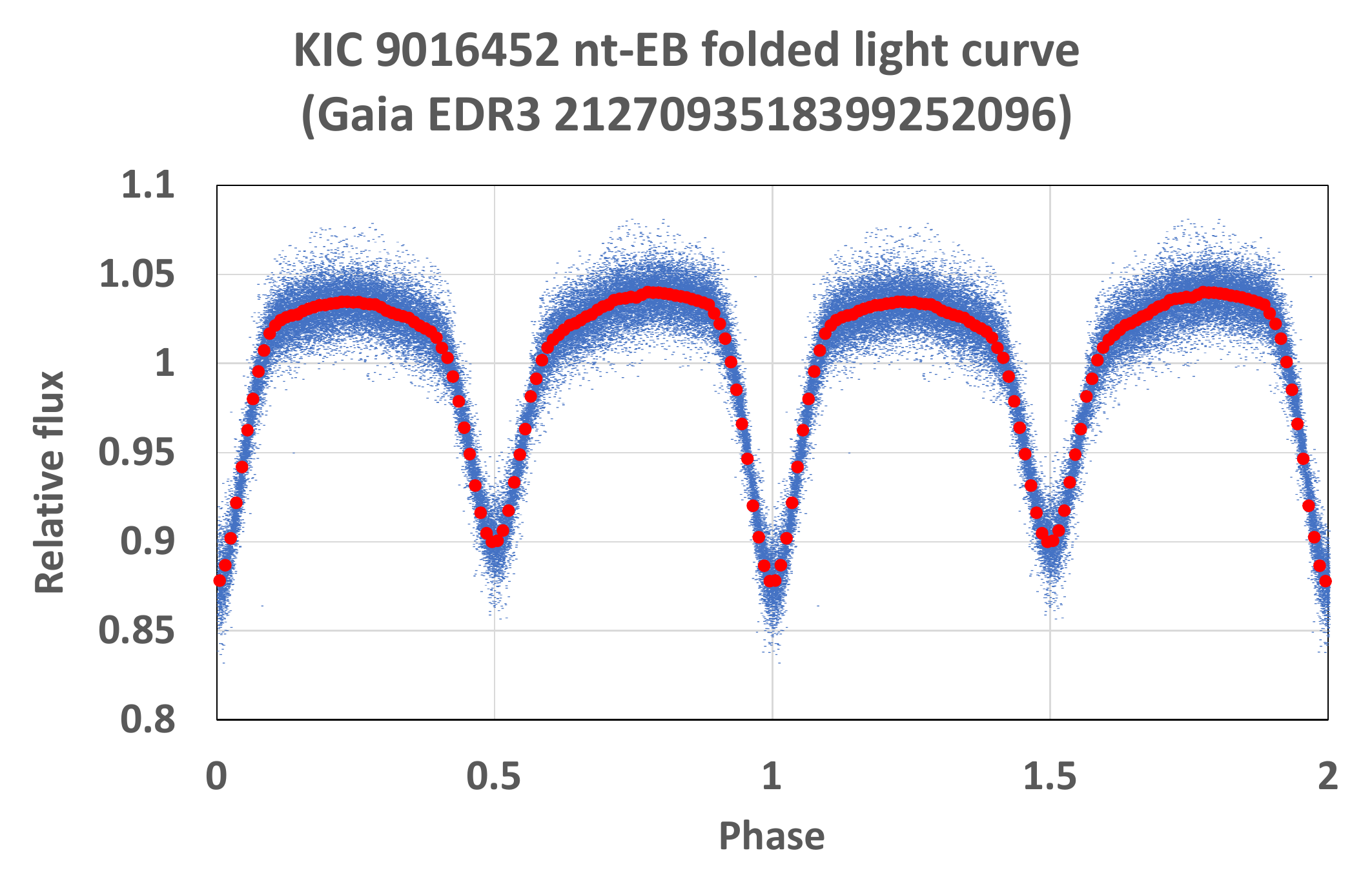}
    \includegraphics[width=\columnwidth]{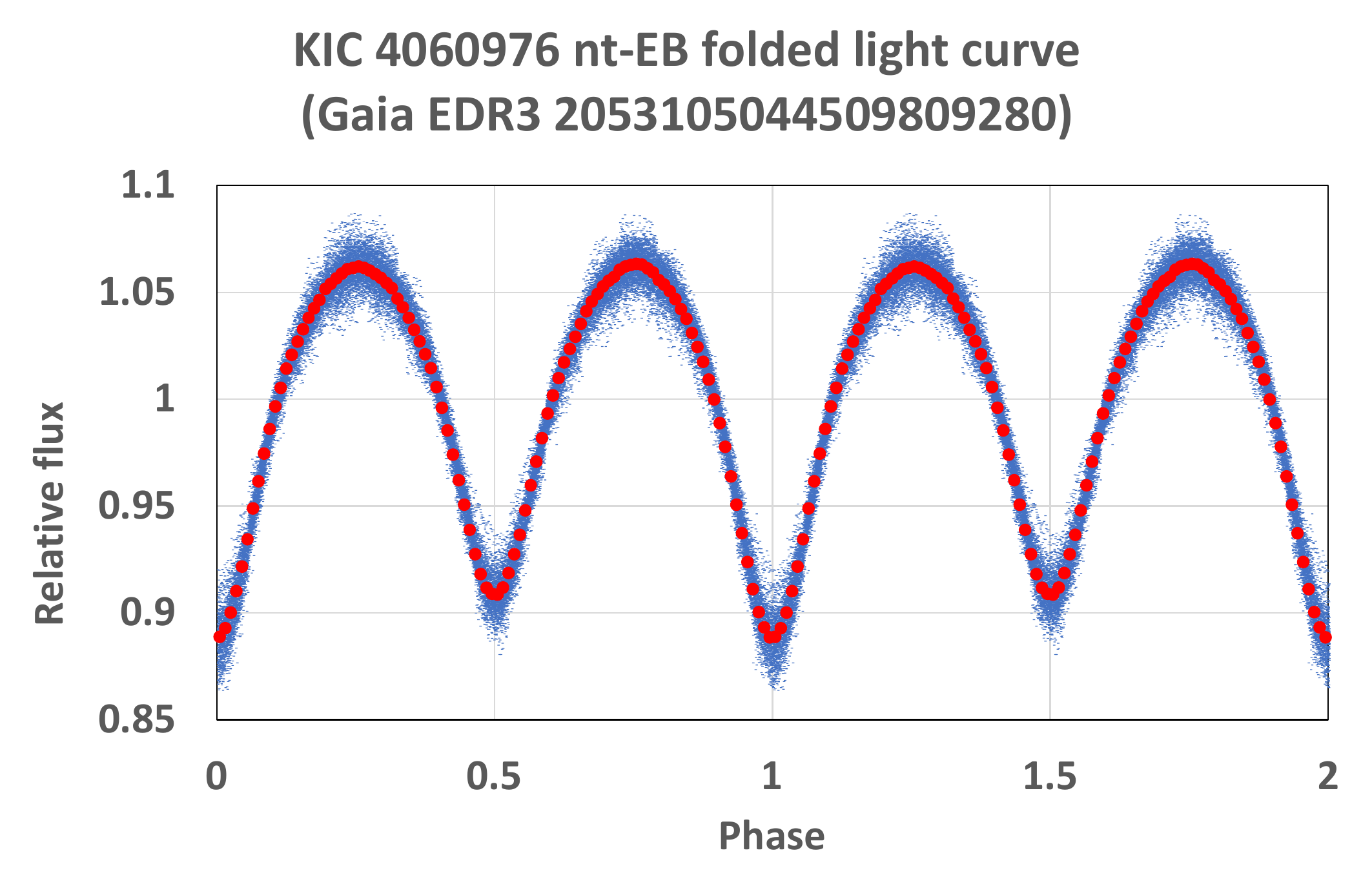}
    \includegraphics[width=\columnwidth]{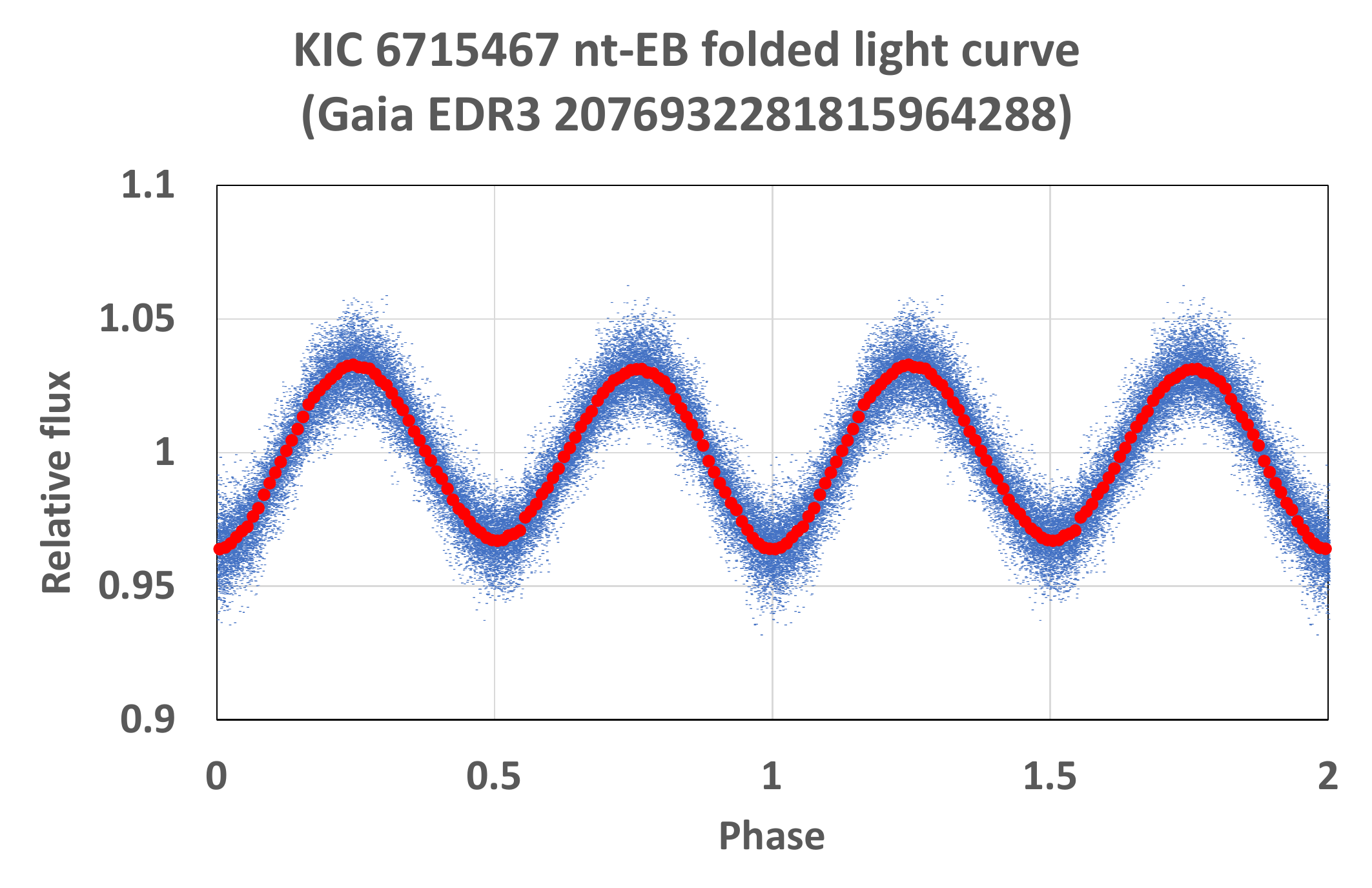}
\caption{Example folded light curves. From the top: EA type, c=0.472; EB type, c=0.667; EW type, c=0.764; ELL type, c=0.971. The charts show the individual observations (\textit{blue}) and the phase bin average fluxes (\textit{red}). The \textit{Gaia} EDR3 candidate sources are shown in brackets.}
\label{fig:folded_lcs}
\end{figure}

A typical example of a phase-folded light curve is shown for each class of eclipsing binary in Fig. \ref{fig:folded_lcs}. Note that outlying data points more than $4\sigma$ from their respective phase bin averages have been discarded. Note also that many candidate light curves are contaminated by nearby relatively bright KIC main objects, and the minima depths are thus unreliable.

\subsection{Identification of source objects} \label{sub:source_id}

\begin{figure}
\centering
    \includegraphics[width=\columnwidth]{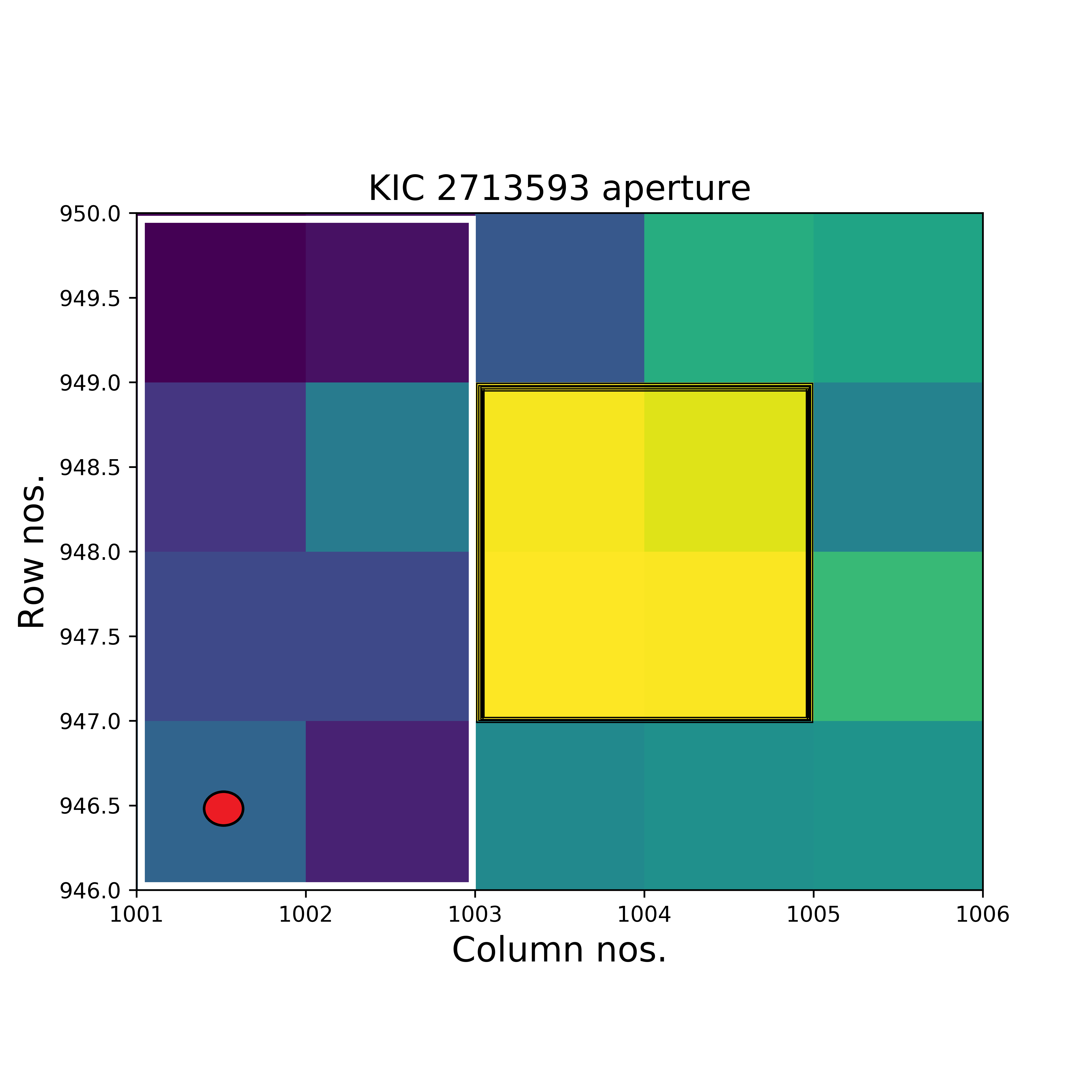}
    \includegraphics[width=\columnwidth]{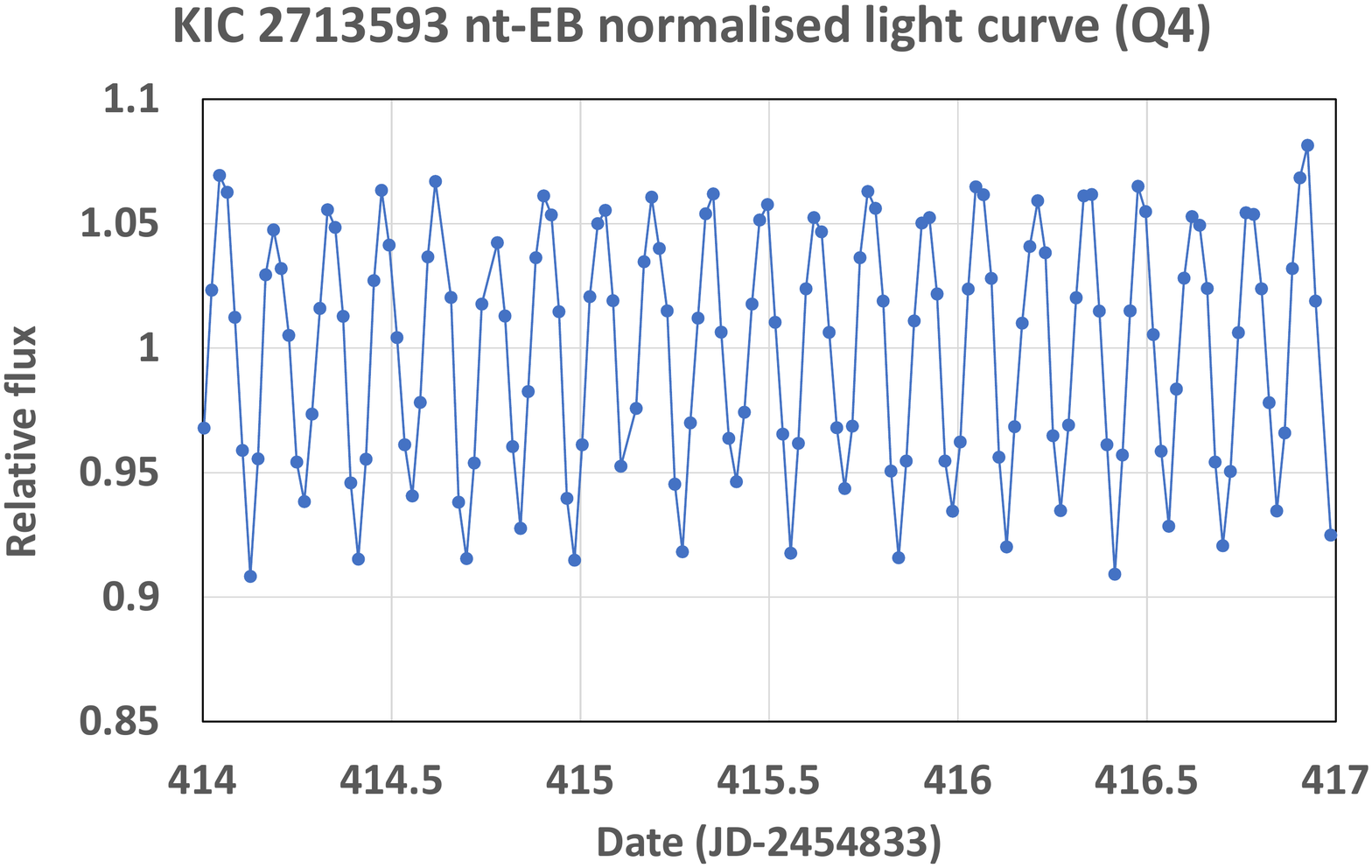}
\caption{KIC 2713593 aperture mask (\textit{top}) with the main eclipsing binary pixel marked with a red circle (source: Lightkurve) and the partial combined and normalised light curve for the eight outlined pixels where the binary signal was detected (\textit{bottom}). The four bright pixels used in the \textit{Kepler} pipeline to extract the main target light curve are outlined in black.}
\label{fig:2713593_pix}
\end{figure}

\begin{figure*}
  \centering
  \includegraphics[width=\textwidth]{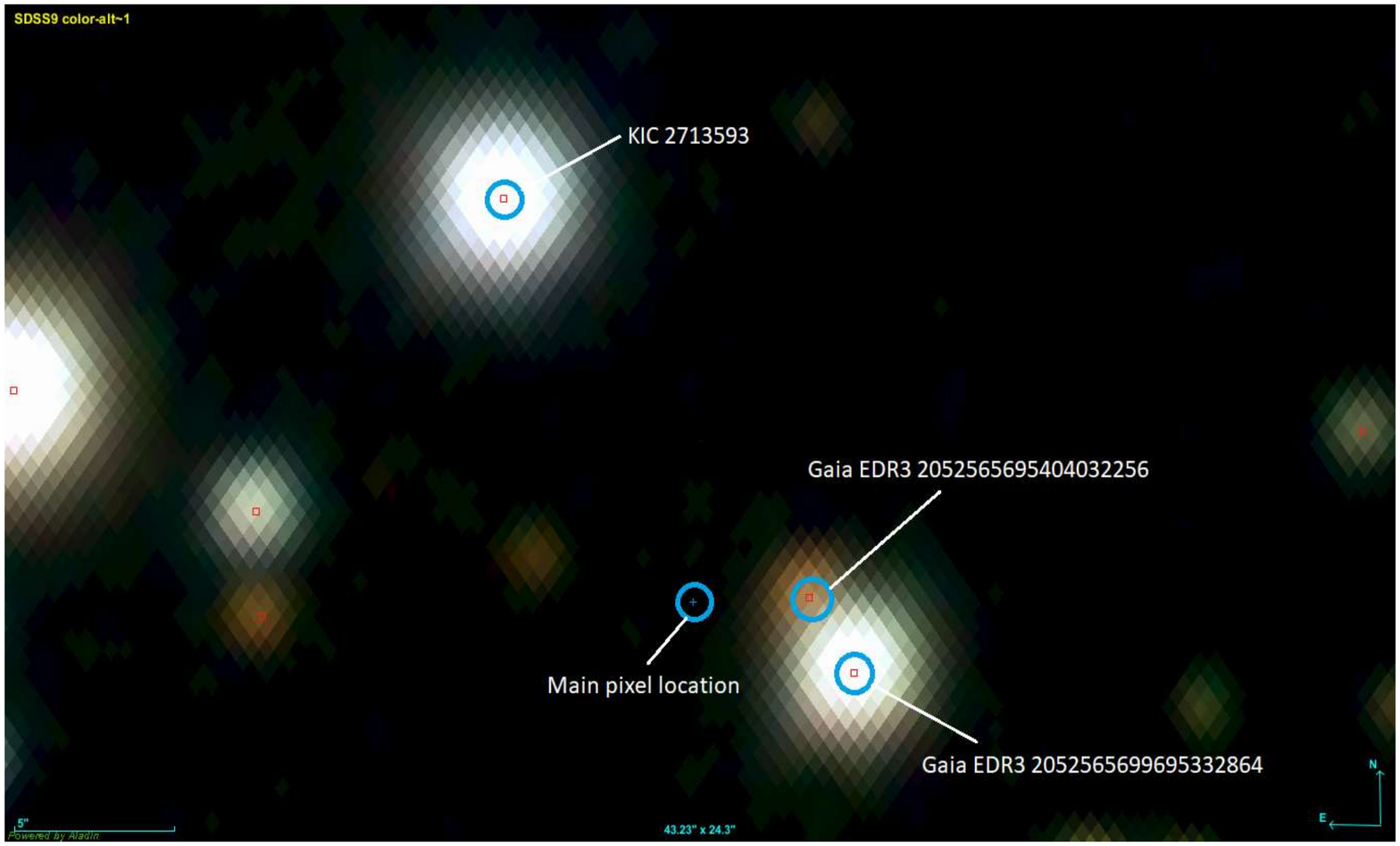}
  \caption{Aladin sky map showing the area around KIC 2713593. The main pixel location for the eclipsing binary and two candidate sources are marked. The FOV is $43.23\arcsec$ across by $24.3\arcsec$ high.}
  \label{fig:2713593_aladin}
\end{figure*}

\begin{figure}
  \centering
  \includegraphics[width=\columnwidth]{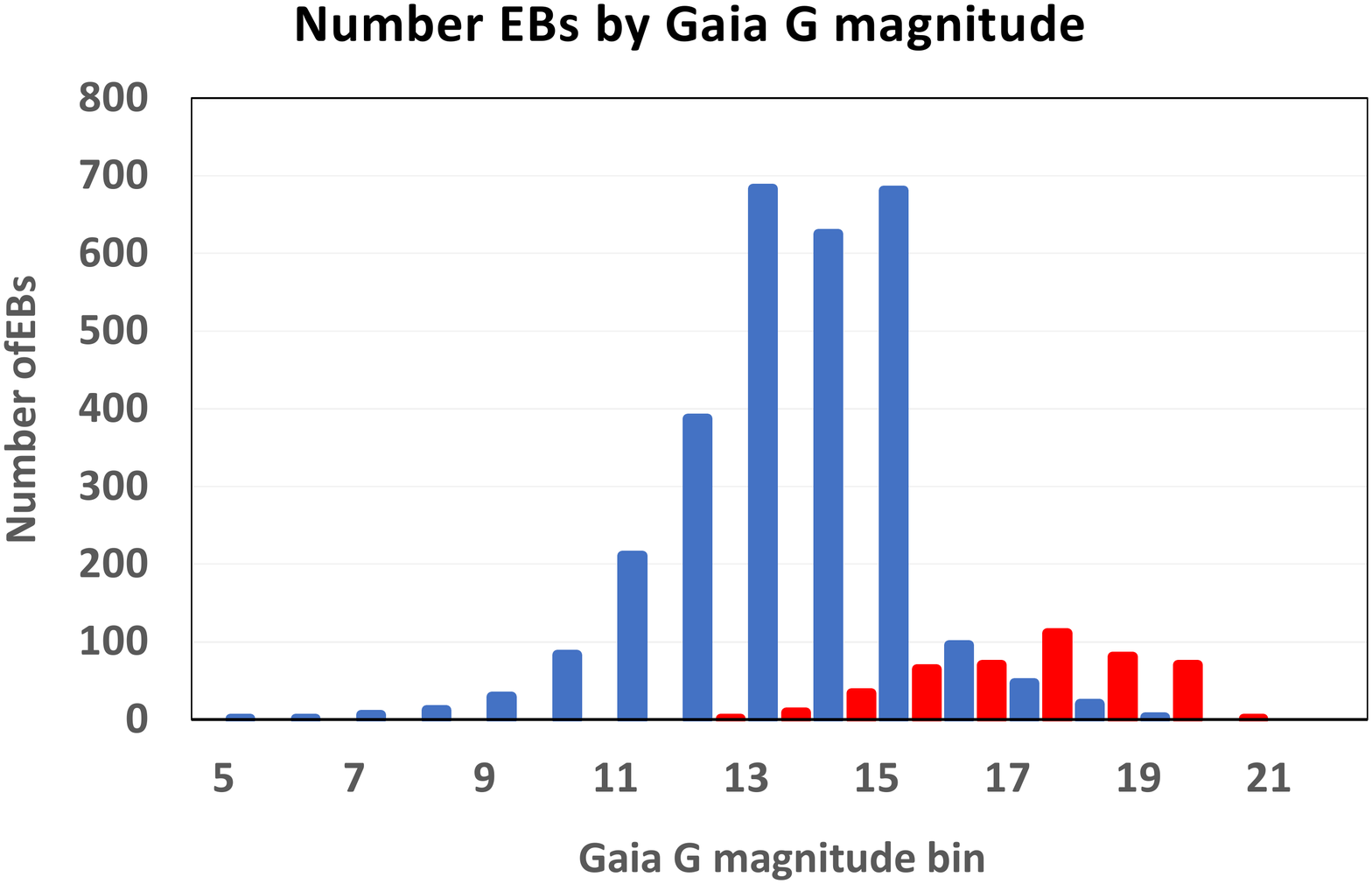}
  \caption{Gaia G magnitude distribution of the \textit{Kepler} Eclipsing Binary Catalogue \citep{kirk2016} (\textit{blue}) and the nt-EBs presented in this work (\textit{red}). Only those nt-EBs with a single candidate source are included. The magnitude axis labels denote the minimum value for each bin.}
  \label{fig:ebs_by_mag}
\end{figure}

The potential sources for the nt-EBs were found from the right ascension (RA) and declination (DEC) of the main pixel, that is, the pixel with the greatest amplitude of flux variation. Candidates for the source objects are those at, or close to, the main pixel coordinates. In the majority of cases, the pixels found lie on the edge of the KIC aperture mask and the centre of the candidate PSF lies outside the mask. As a result, in many cases, there are several possible source candidates for a single SPEB. An example is shown in Figs.~\ref{fig:2713593_pix} and \ref{fig:2713593_aladin}. Fig.~\ref{fig:2713593_pix} shows the pixel-by-pixel aperture mask for KIC 2713593 (top) with the red circle marking the main pixel where the nt-EB signal was detected. The four bright pixels used in the \textit{Kepler} pipeline to extract the main target light curve are outlined in black. Eight pixels in the aperture were detected with the same signal as the main nt-EB pixel (outlined in white in the figure). Part of the combined and normalised light curve for these eight pixels is then shown below. The coordinates for the main pixel are obtained using Astropy and Fig.~\ref{fig:2713593_aladin}, obtained from Aladin, shows the area of sky around this location, with the pixel location, KIC 2713593 and the two possible source candidates marked. The two candidates are relatively faint having Gaia G magnitudes of 19.70 (Gaia EDR3 2052565695404032256) and 16.10 (Gaia EDR3 2052565699695332864). The candidates found are generally fainter than those in the current \textit{Kepler} eclipsing binary catalogue. This is illustrated in Fig.~\ref{fig:ebs_by_mag} showing the distribution of the cat-EBs and nt-EBs by magnitude. The cat-EBs have an average Gaia G magnitude $\approx 14.0$ whereas the average magnitude for the nt-EBs (which have a single candidate source only) $\approx 18.2$.

\begin{figure}
  \centering
  \includegraphics[width=\columnwidth]{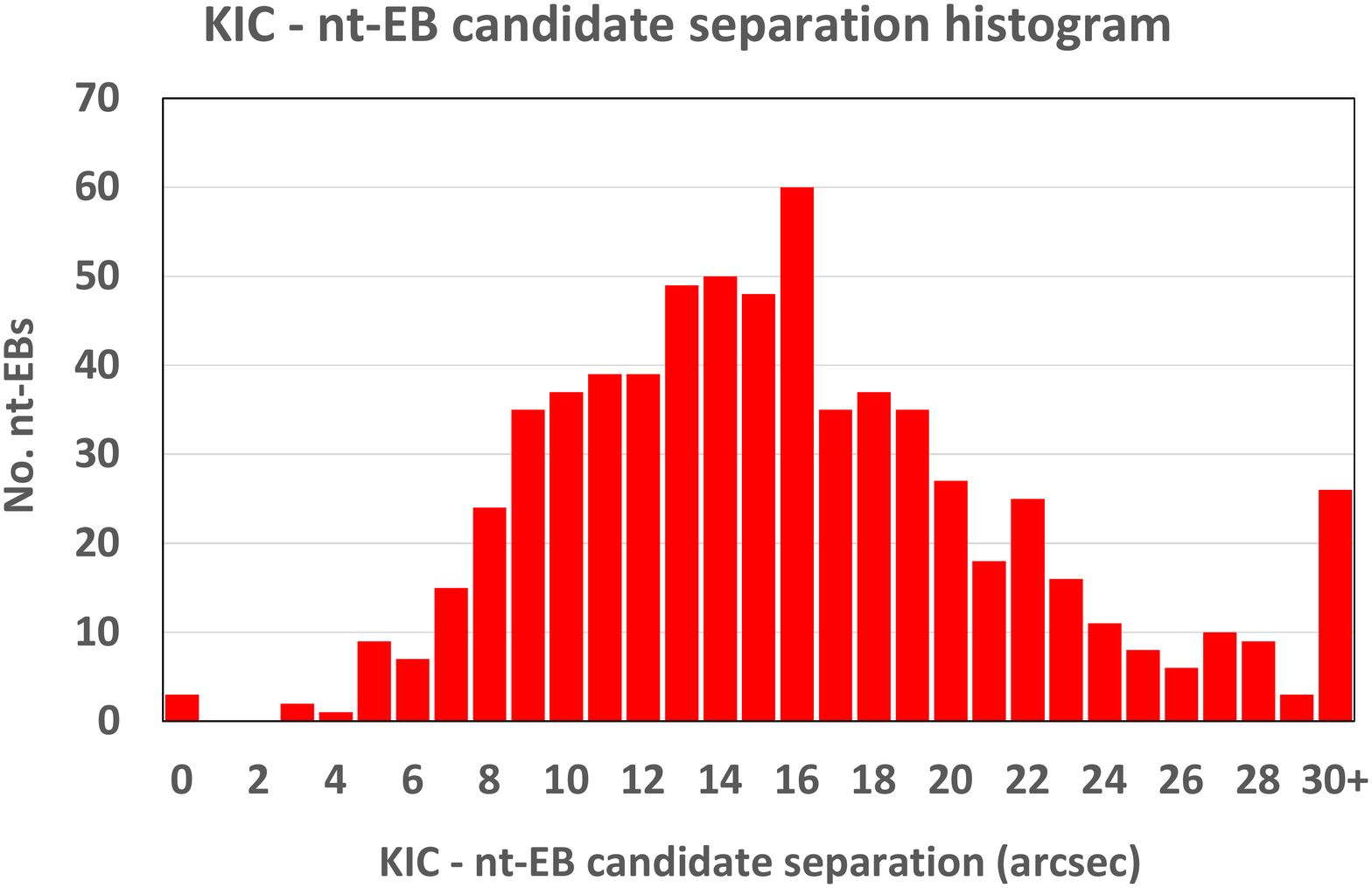}
  \caption{Distribution of the separation (in arcsec) between KIC main targets and the candidate SPEBs found in their backgrounds. The separation axis labels denote the minimum value for each bin.}
  \label{fig:KIC_SPEB_sep}
\end{figure}

The average separation of the KIC main targets from the nt-EBs found in their apertures $\approx 17\arcsec$. Fig.~\ref{fig:KIC_SPEB_sep} shows the distribution of these separations, with all except 26 nt-EBs lying within $30\arcsec$ of their respective KIC main targets. The distribution shows a peak at $\sim 16\arcsec$ separation followed by a decline, presumably due to a combination of increasing area searched with increasing radius from the main target and a decline in the number found as the separation increases beyond the peak and and nt-EB PSFs are only detected in the larger main target apertures.

Note that in the three cases that appear in the $0-1\arcsec$ separation bin, the main KIC target is itself the SPEB candidate, these being KICs 5982499, 10156116 and 10670920.

In nine instances, no candidates were found close to the main pixel coordinates, and, since these candidates also generally exhibited poor quality light curves and uncertain morphologies, they were deemed to be false positives and discarded. The identified Gaia candidates for the eclipsing binaries are listed in Table \ref{tab:results}. 

\section{Results} \label{sec:results1}
The full set of results, including classifications, periods, and candidate source objects are shown in Table \ref{tab:results} in the Appendix. Note that the KIC values in the table refer to the KIC aperture masks in which the candidates were found, not the KIC main targets themselves (except for the three cases listed above). Of the 547 objects listed, the majority, 446, are W UMa types, while 80 ellipsoidal, 14 $\beta$ Lyrae  and 7 Algol types have also been found.

\subsection{Period Distribution} \label{sub:period_dist}
\begin{figure}
  \centering
  \includegraphics[width=\columnwidth]{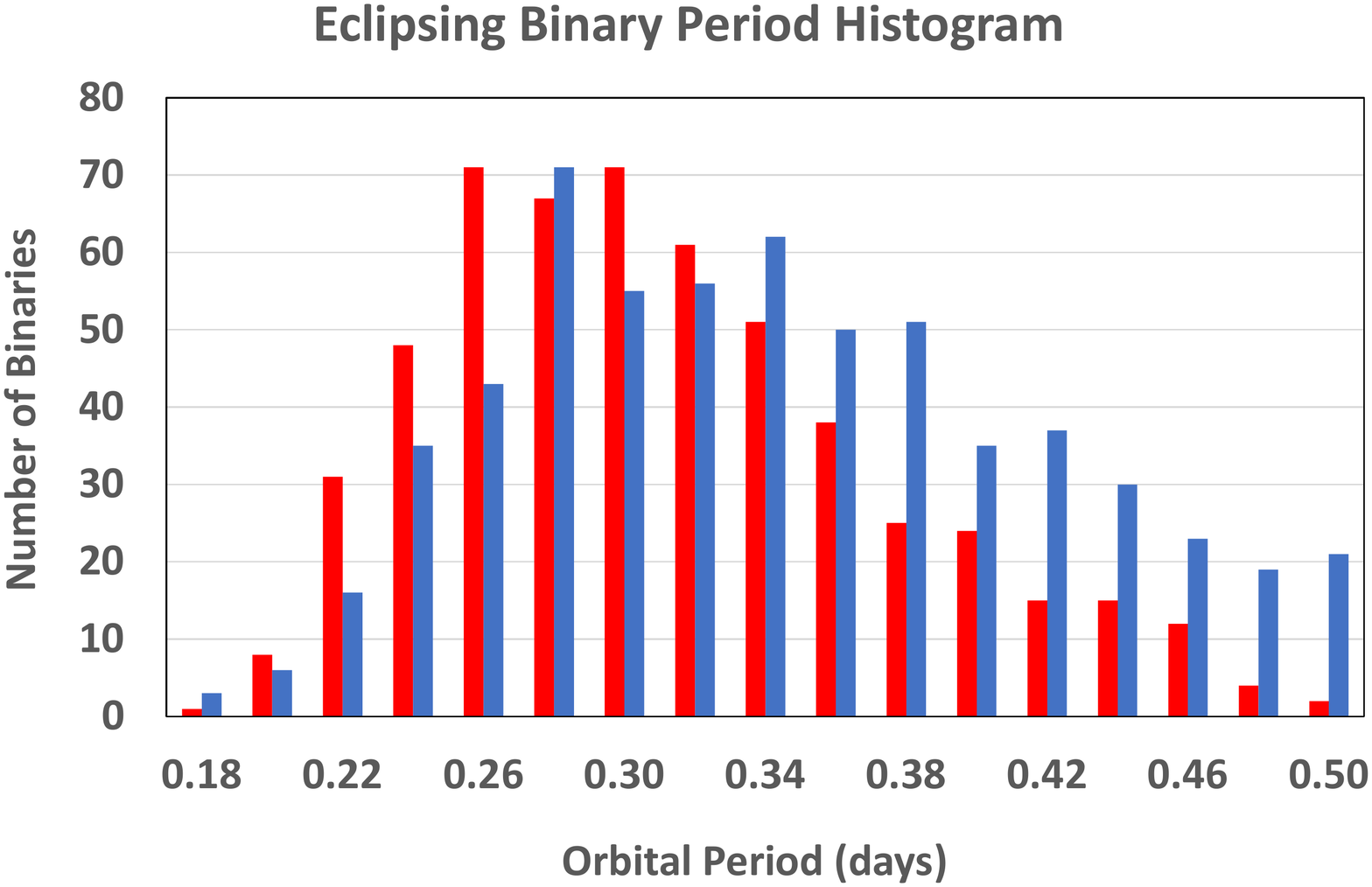}
  \caption{Period histogram for eclipsing binaries. The nt-EBs presented in this paper are shown in \textit{red} and the equivalent set from the \textit{Kepler} EB catalogue are shown in \textit{blue}. The orbital period axis labels denote the minimum value for each bin.}
  \label{fig:SPEB_period_histo}
\end{figure}

 Fig.~\ref{fig:SPEB_period_histo} shows the orbital period distribution for 544 of the 547 nt-EBs presented in this work (\textit{red}) and the equivalent set from the \textit{Kepler} EB catalogue (\textit{blue}). Note that the three nt-EBs with periods $\sim 0.9$ days have been omitted from the histogram. The distribution follows the well-known form with a peak at around 0.29 days, and a cutoff at around 0.22 days below which few eclipsing binaries are observed \citep{rucinski1992}.
 
 It is noticeable that the nt-EB distribution is skewed towards the short period end by comparison with the cat-EBs. It is assumed that this is due to the pre-selection of KIC target stars by spectral type (F, G and K only) and magnitude, thus biasing the EB selection away from short periods.
 
Calculations by \citet{Stepien2006} suggest that the cutoff is related to the evolution of low mass binaries whereby orbital decay is driven by magnetic wind-driven angular momentum loss. A binary will change from detached to contact type when the primary star overflows its Roche lobe, but the time taken to reach this point increases with decreasing stellar mass. For a primary of 1 M\textsubscript{$\odot$}, this will be at an age of approximately 7.5 Gyr (depending on the primary/secondary mass ratio) when the orbital period is approximately 0.4 days. Such stars then coalesce before reaching the 0.22 day cutoff period. \citet{Stepien2006} showed that the short period limit of 0.22 days corresponds with a lower total binary mass limit of 1.1 -- 1.2 M\textsubscript{$\odot$}, and masses lower than this will require a time greater than the age of the universe to evolve into a contact binary. 

However, it appears that the 0.22 day period cutoff is less strict than previously thought. Eight W UMa type binaries are identified in this work with periods less than 0.22 days, and with one having a period of 0.185508 days (the nt-EB in the background of KIC 4481056). Other recent papers have also reported very short period binaries with $P < 0.22$ days, notably \citet{soszynski2015ultrashortperiod} who report 242 such binaries found in the OGLE survey.

\subsection{Spatial Distribution} \label{sub:spatial_dist}

\begin{figure}
\centering
    \includegraphics[width=\columnwidth]{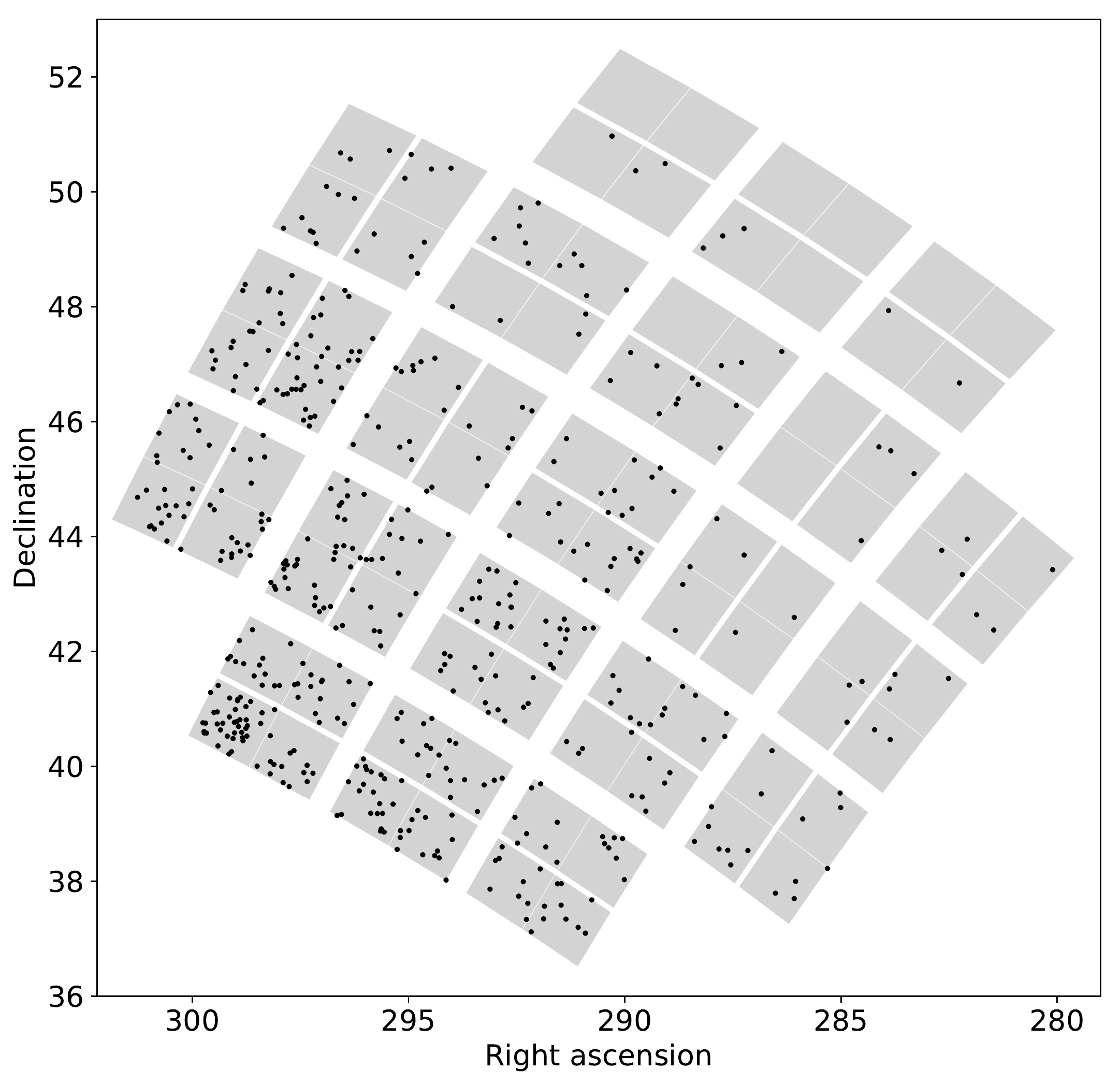}
    \includegraphics[width=\columnwidth]{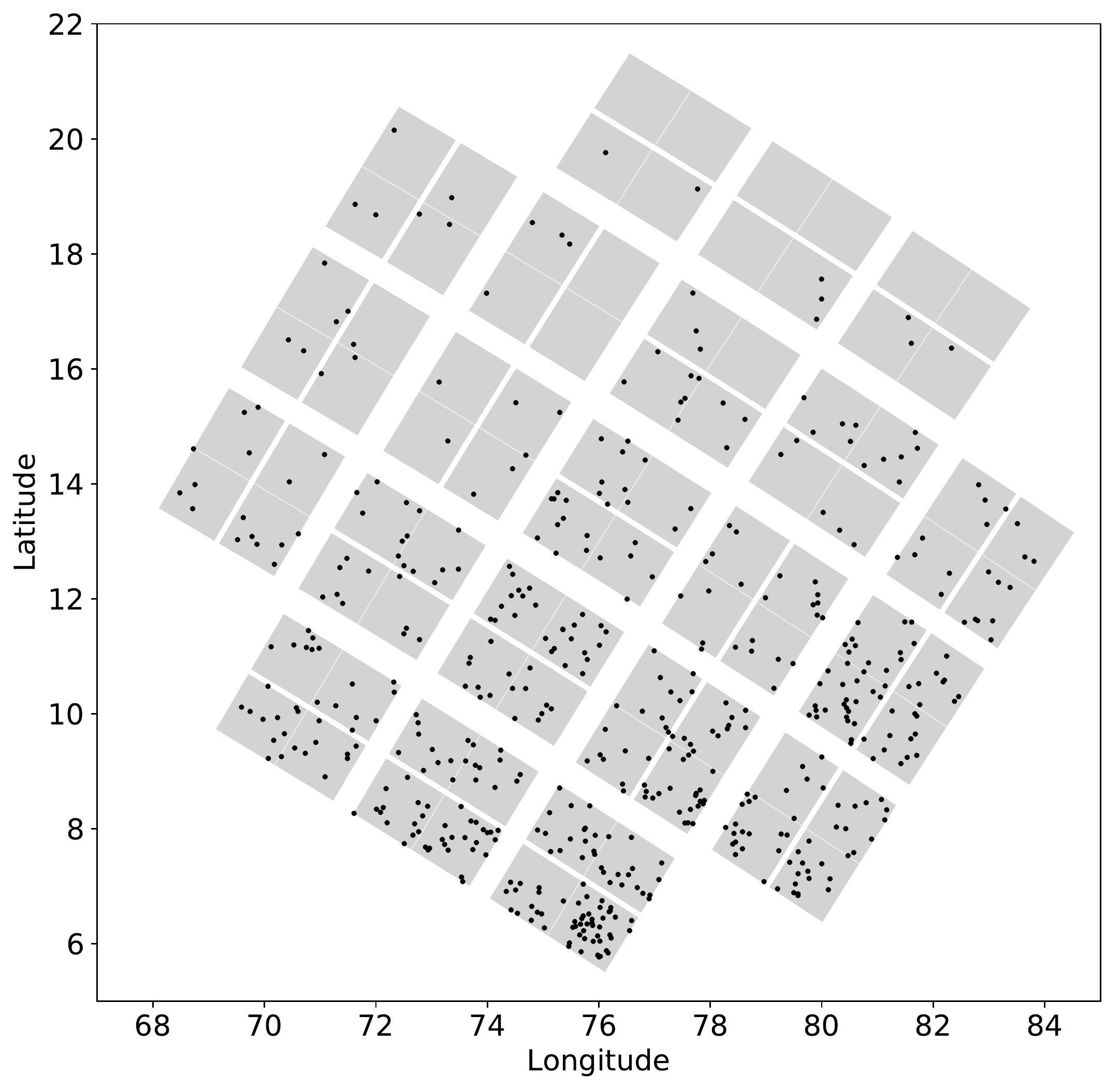}
\caption{nt-EB spatial distribution by right ascension and declination (\textit{top}) and galactic latitude and longitude (\textit{bottom}).} 
\label{fig:radec_gal}
\end{figure}

Fig.~\ref{fig:radec_gal} shows the RA and DEC distribution of the newly found nt-EBs across the \textit{Kepler} field of view (\textit{top}) and also the galactic latitude and longitude distribution (\textit{bottom}).

\begin{figure}
  \centering
  \includegraphics[width=\columnwidth]{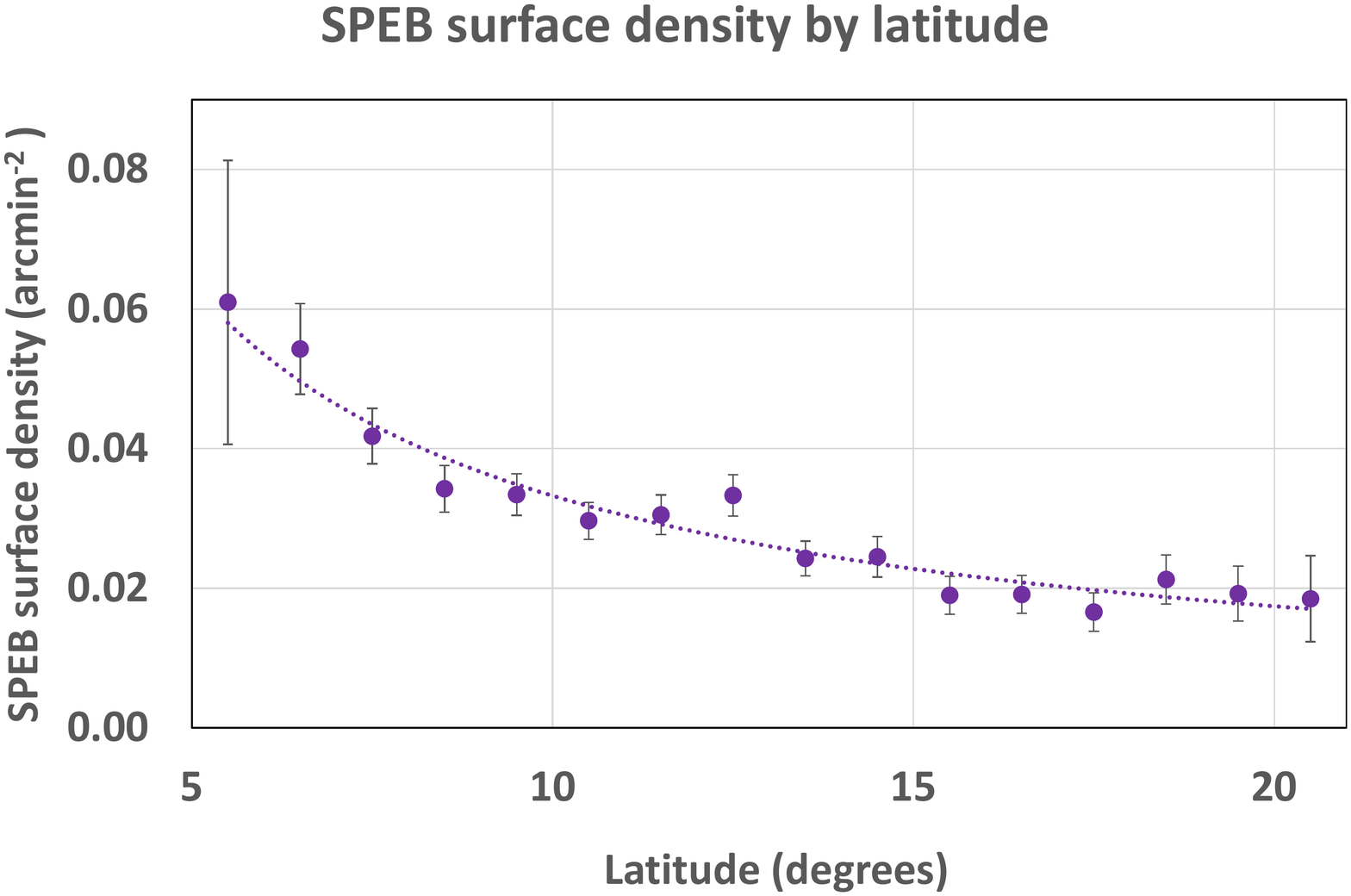}
  \includegraphics[width=\columnwidth]{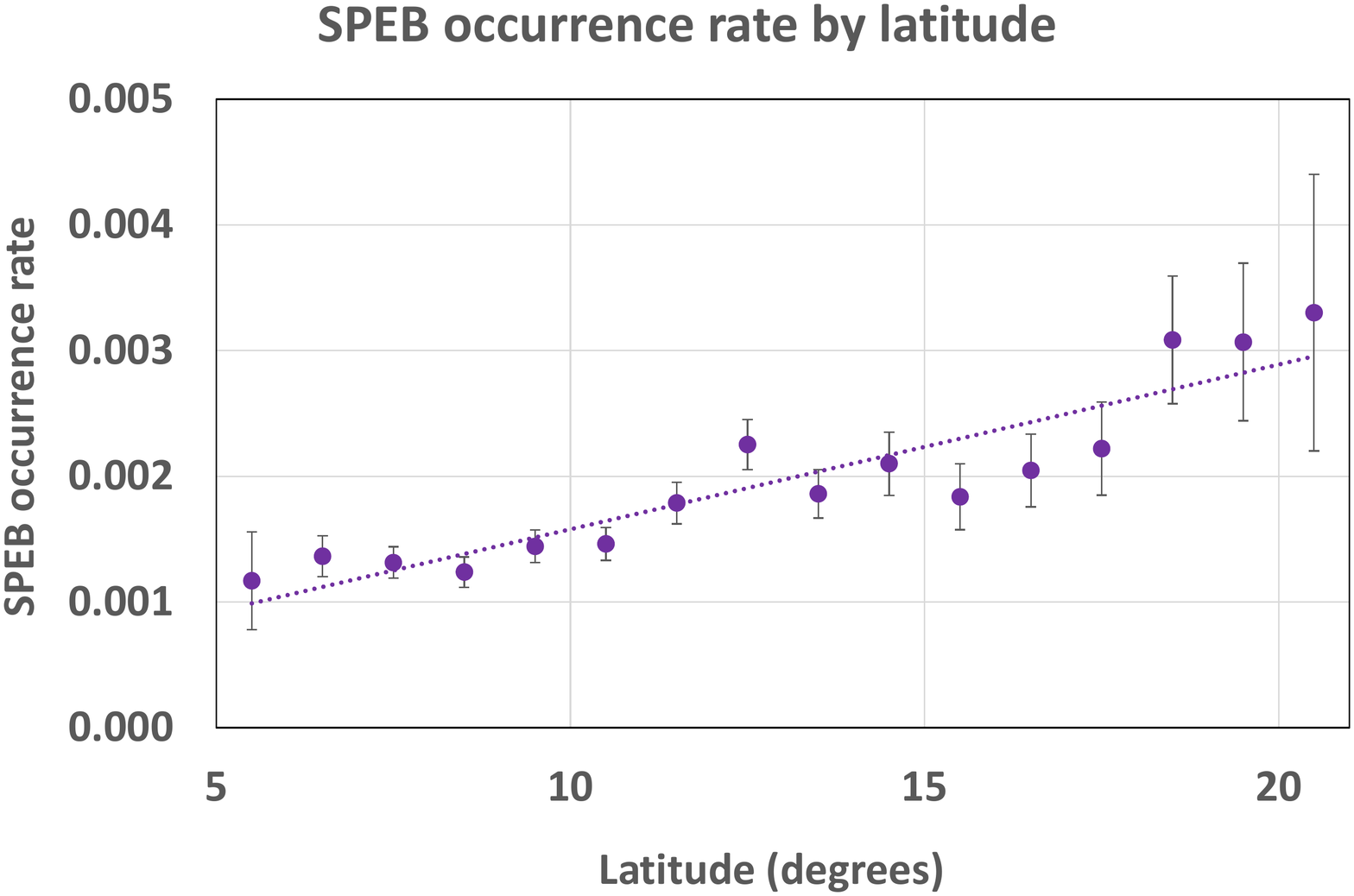}  
  \caption{\textit{Top}: Variation of SPEB surface density by galactic latitude for the combined nt- and cat-EBs in the same period range. The dotted line is a power law fit to the surface density, $\rho$, and is given by $\textrm{log}(\rho) = (-0.931 \pm 0.070)\textrm{log}(b) - (0.547 \pm 0.077)  \textrm{ arcmin}^{-2}$ where \textit{b} is the galactic latitude in degrees. \textit{Bottom}: SPEB occurrence rate for the combined nt- and cat-EBs.}
  \label{fig:speb_density}
\end{figure}

It can be seen from Fig.~\ref{fig:radec_gal} that the number of nt-EBs declines with increasing galactic latitude. This is illustrated more clearly in Fig.~\ref{fig:speb_density} which shows the spatial variation in two different ways.  For this purpose, the nt-EBs have been combined with the cat-EBs in the same period range to provide a more comprehensive view of the SPEB population. Firstly, in the \textit{top} chart, the variation of surface density with latitude is shown for the combined set of SPEBs. Surface density is defined to be $N_{EB}/A_{KIC}$ where $N_{EB}$ is the number of SPEBs in each latitude bin (each bin being 1 degree of latitude wide) and $A_{KIC}$ is the total area (in arcmin$^{2}$) of the KIC target aperture masks searched in each latitude bin. Since most of the nt-EB candidates lie outside of, but close to, the aperture masks, the effective aperture mask area searched has been assumed to include one pixel width around each mask. This results in a total of $9.176 \times 10^{6}$ pixels across approximately 170,000 KIC aperture masks. The actual aperture masks (without the extra one pixel width) averaged 29.2 pixels per KIC, but for the KIC masks on which nt-EBs were found, the average actual number of pixels is greater at 47.5, presumably because the chances of finding a background binary increase with aperture size. The uncertainty in each surface density value, following Poisson statistics, is calculated as $\sqrt{N_{EB}}/A_{KIC}$.

The surface density, $\rho$, is well fitted by a simple power law given by:
\begin{equation}
\textrm{log}(\rho) = (-0.931 \pm 0.070)\textrm{log}(b) - (0.547 \pm 0.077)  \textrm{ arcmin}^{-2}
\end{equation}
where \textit{b} is the galactic latitude in degrees. This decline is attributed, at least in part, to the decline of stellar density with increasing latitude/height above the galactic plane, and thus a lower average number of background stars per KIC aperture mask. 

Secondly, the effect of declining stellar density has been removed by estimating the SPEB occurrence rate by latitude, defined to be the fraction of examined stars that are SPEBs, i.e. $N_{EB}/N_{*}$ where $N_{EB}$ is the number of SPEBs in a latitude bin and $N_{*}$ is the number of stars examined in the same bin, that is, the number of stars within the effective aperture masks of the KIC objects in the bin.

$N_{*}$ is estimated by counting the number of Gaia EDR3 objects within a given radius of each KIC target object, either $30\arcsec$ or $1\arcmin$ depending on the density of stars and the number of KICs in a bin. The resultant count is then scaled to the effective aperture mask area of the KIC, and the counts for the KICs summed to give the estimated number of stars examined in each bin. Again, the uncertainty in each occurrence rate value follows Poisson statistics and is calculated as $\sqrt{N_{EB}}/N_{*}$.

The result, shown in the \textit{lower} chart of Fig.~\ref{fig:speb_density}, indicates an approximately linear rising trend in the occurrence rate with latitude. This is unexpected when compared with the  exponential decline in the occurrence rate noted by \citet{kirk2016}. However, the results cannot be compared directly, since the analysis by \citet{kirk2016} includes \textit{Kepler} main target eclipsing binaries of all periods resulting from a convoluted pre-selection process (principally by spectral type and log $g$ value) designed to maximize planet yields. Nonetheless, it does appear that the latitude variation of short period binaries differs from that of longer period binaries, possibly reflecting the changing stellar age profile, with stars at higher latitudes being on average older than those at low latitudes.

\subsection{Physical and Geometric Parameter Estimates}

In the absence of spectroscopic observations, it is difficult to obtain reliable estimates of stellar mass and radius and orbital parameters for the source candidates obtained from the \textit{Gaia} catalogue. However, \citet{Gazeas_Stepien2008} provide calculations, based on correlations obtained from observations, to approximately estimate some of these parameters for W UMa type binaries, solely on the basis of orbital period.

In particular, they obtain relations for binary absolute magnitude, $M_V$, component masses, $M_1$ and $M_2$, and orbital semi-major axis, $a$, as follows.

Combining period-colour and period-magnitude-colour relations leads to a magnitude-period relation (period in days):
\begin{equation}
M_V = -8.4\log(P)+0.31
\end{equation}

Calibrating observed stellar masses and periods leads to the following mass-period relations with an estimated accuracy of 15\% (masses in solar units):
\begin{equation}
\log(M_1) = (0.755 \pm 0.059)\log(P) + (0.416 \pm 0.024)
\end{equation}
\begin{equation}
\log(M_2) = (0.352 \pm 0.166)\log(P) - (0.262 \pm 0.067)
\end{equation}

Kepler's third law then leads to the expression for the semi-major axis, $a$ (in solar radii):
\begin{equation}
P = 0.1159a^{3/2}(M_1+M_2)^{1/2}
\end{equation}

The critical Roche lobe radii, $R_1$ and $R_2$ (in solar radii), may then be estimated from the semi-major axis and mass ratio, $q$ (i.e. $M_2/M_1$), using the expressions derived from \citet{Eggleton1983}:
\begin{equation}
\frac {R_1}{a} = \frac {0.49q^{-2/3}}{0.6q^{-2/3}+\ln(1+q^{-1/3})}
\end{equation}
\begin{equation}
\frac {R_2}{a} = \frac {0.49q^{2/3}}{0.6q^{2/3}+\ln(1+q^{1/3})}
\end{equation}

The resulting parameters for the 446 W UMa type binaries are shown in Table \ref{tab:phys_param} and a summary of the results, showing the average, minimum and maximum values found for each parameter is shown in Table \ref{tab:phys_summ}. It is interesting that the minimum primary mass, 0.730 M\textsubscript{$\odot$}, (from the nt-EB in the background of KIC 4481056 with period 0.185508 days) corresponds closely with the limit of $0.7$ M\textsubscript{$\odot$} calculated by \citet{Stepien2006} which should require a time greater than the age of the universe to evolve into a contact binary. However, \citet{Jiang2015} propose that dynamical instability of mass transfer could greatly reduce the time take for a detached binary to evolve into a contact binary and reduce the minimum orbital period to approximately $0.165$ days and the minimum observed primary mass to $0.63$ M\textsubscript{$\odot$}. Moreover, even shorter orbital periods in M-type dwarf binaries, as low as $0.112$ days, have been reported by \citet{Nefs2012}, suggesting that further work is required on the evolution of contact binaries.

Since the results we present in Table \ref{tab:phys_param} are only estimates based on the statistics of short period eclipsing binaries, the values obtained are only usable as initial inputs to further fitting processes. However, this work is outside the scope of this paper.

\begin{table}
\centering
\begin{tabular}{ |c|c|c|c| }
\hline
 Parameter & Average & Minimum & Maximum\\
 \hline
 $M_{V}$ (\rm{mag}) & 4.544 & 2.899 & 6.456\\
 $M_{1}$ $(\textit{M}_{\odot})$ & 1.096 & 0.730 & 1.528\\
 $M_{2}$ $(\textit{M}_{\odot})$ & 0.364 & 0.302 & 0.427\\
 a $(\textit{R}_{\odot})$ & 1.721 & 1.354 & 2.100\\
 q $(\textit{M}_{2}/\textit{M}_{1})$ & 0.336 & 0.279 & 0.414\\
 $R_{1}$ $(\textit{R}_{\odot})$ & 0.819 & 0.618 & 1.033\\
 $R_{2}$ $(\textit{R}_{\odot})$ & 0.497 & 0.414 & 0.579\\
 \hline
\end{tabular}
\caption{Summary of physical parameter calculations. See text for details.}
\label{tab:phys_summ}
\end{table}

\subsection{Eclipse Timing Variations} \label{sub:ETV}

The long-term nature and precision of the \textit{Kepler} photometric observations enables analysis of the changes in the orbital periods of the systems in the way that was employed for the primary targets in the Kepler FOV  \citep{Balaji2015,Borkovits2015,Borkovits2016,Conroy2014}. The period changes of eclipsing binaries are traditionally examined by eclipse timing variation (ETV) analysis via $O-C$ (Observed -- Calculated) diagrams. 

The $O-C$ diagrams can be used to reveal period changes caused, for example, by the periodically varying distance between the system and the observer due to the gravitational effect of a third object. This is the so-called light-travel-time effect (LTTE; see e.g. \citealt{Hajdu19}). The  mathematical formula of LTTE  was first described by \citet{Irwin1952} in the following form:

\begin{equation}
\label{LTTEfunction}
\Delta_\mathrm{LTTE}=-\frac{a_\mathrm{AB}\sin i_2}{c}\frac{\left(1-e_2^2\right)\sin\left(v_2+\omega_2\right)}{1+e_2\cos v_2},
\end{equation}
where $a_\mathrm{AB}$ , $i_2$, $e_2$, $\omega_2$ stand  for the semi-major axis of the EB's center of mass around the center of mass of the triple system, the inclination, the eccentricity, and the argument of periastron of the relative outer orbit, respectively.  Furthermore, $c$ is the speed of light and $v_2$ is the true anomaly of the third component.

\begin{figure}
    \centering
    \includegraphics[width=\columnwidth,trim={0 0 0cm 0},clip]{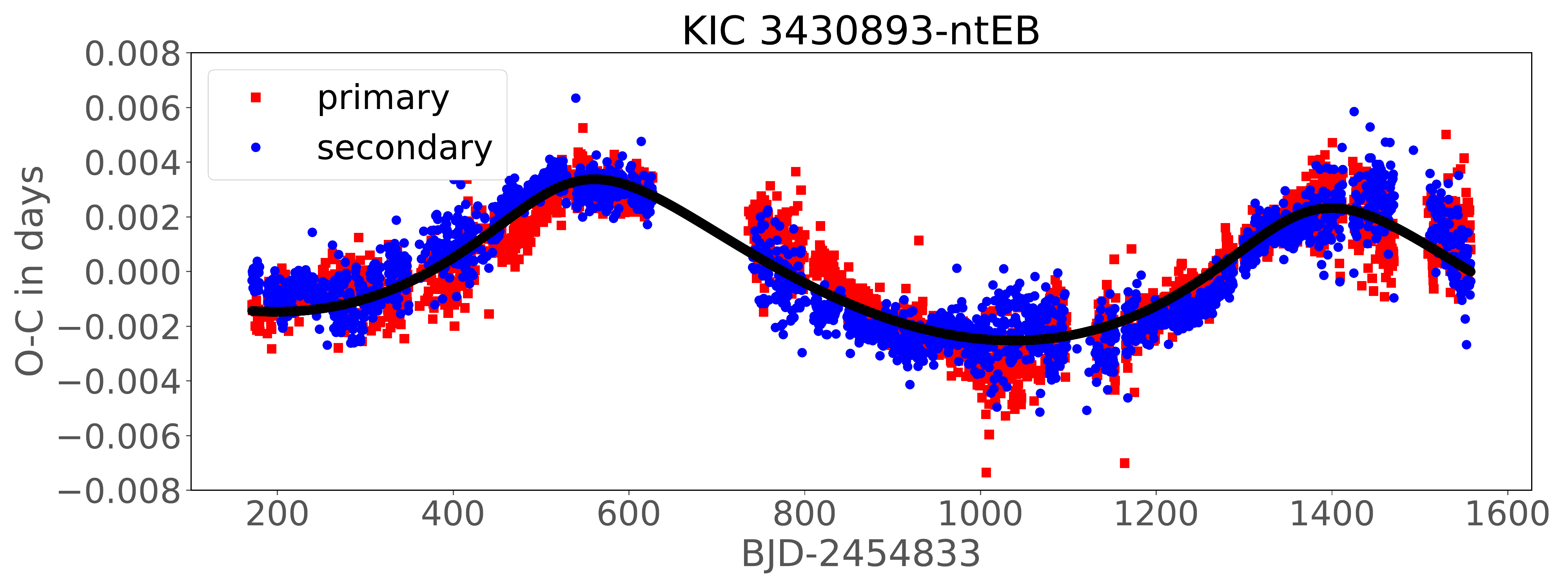}
    \includegraphics[width=\columnwidth,trim={0 0 0cm 0},clip]{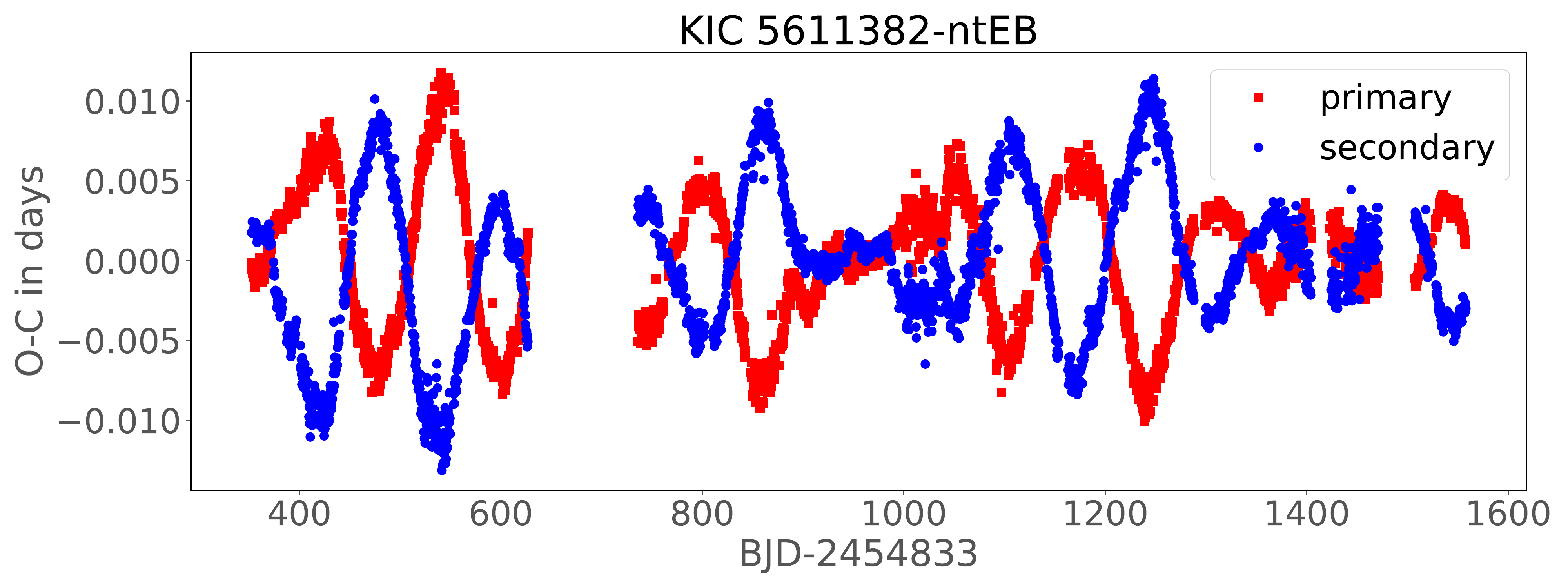}
    \includegraphics[width=\columnwidth,trim={0 0 0cm 0},clip]{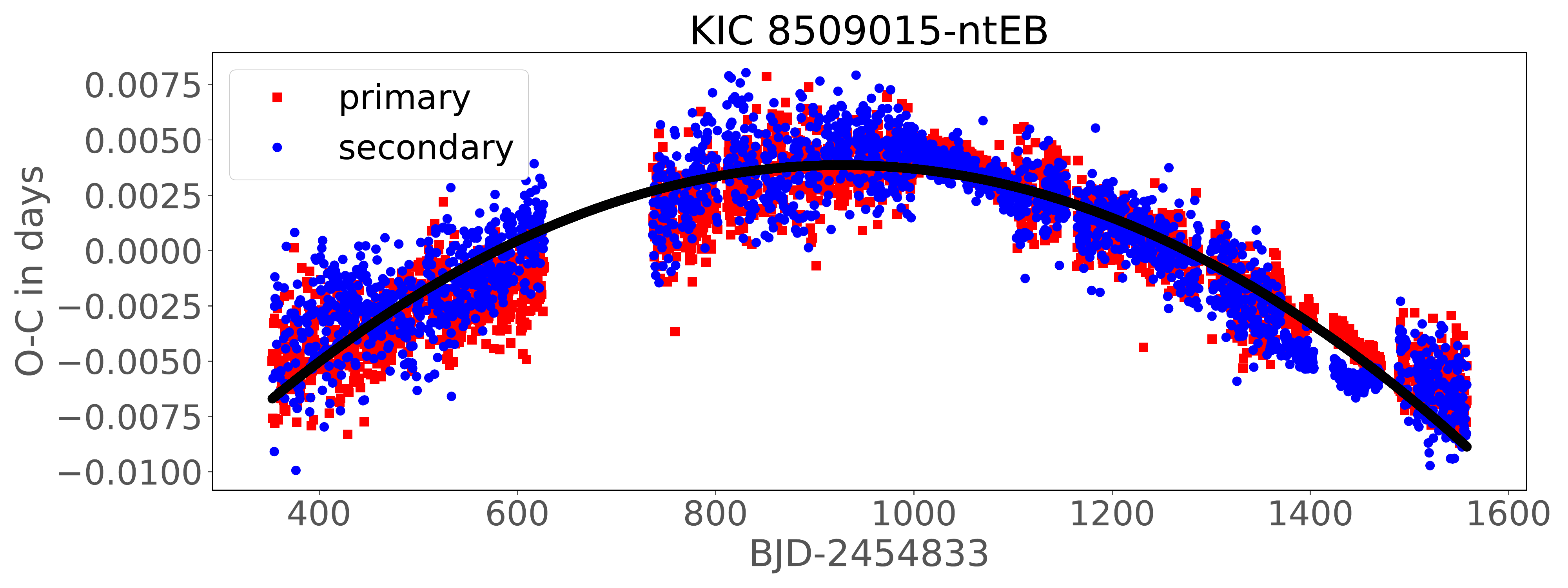}
    \caption{Examples of three close binary systems with typical eclipse timing variations. The red and blue points correspond to the ETV of the primary and secondary eclipses, respectively. The \textit{top} panel presents the O--C of a hierarchical triple system candidate. The \textit{middle} panel shows an anticorrelated quasi periodic variation caused by stellar activity and the \textit{bottom} panel shows a parabolic trend probably caused by mass transfer between the components.}
\label{fig:etv}
\end{figure}

A good example of the LTTE from our EBs is Gaia EDR3 2099555012732795904 (in the background of KIC 3430893), the ETV of which (top panel of Fig.~\ref{fig:etv}) shows a sinusoidal variation with period $P_2 \approx829$ days. We fitted the orbital parameters in Eq. \ref{LTTEfunction} along with the periastron passage time ($\tau_2$) of the third component using the Markov Chain Monte Carlo method  implemented in the \texttt{emcee} Python code \citep{emcee}. The results are listed in Table \ref{tab:ETV_param}.
\begin{table}[]
    \centering

    \begin{tabular}{c c  c}
        \hline
        $P_1$ & days & 0.3706\\\
        $a_{AB}\sin(i_2)$ & $R_{\odot}$ & $100.11^{+0.59}_{-0.50}$ \\
        $P_2$ & days & $828.82^{+2.02}_{-1.66}$\\
        $e_2$ & & $0.37^{+0.01}_{-0.01}$ \\
        $\omega_2$ & rad & $4.27^{+0.03}_{-0.03}$\\
        $\tau_2$ & BJD-2454833 days & $529.41^{+3.22}_{-3.32}$\\
        \hline
    \end{tabular}
    \caption{The fitted LTTE parameters of Gaia EDR3 2099555012732795904.}
    \label{tab:ETV_param}
\end{table}

ETVs may also be affected by stellar spot migration, in which case the ETV curves produced for both the primary and secondary eclipses show anticorrelated quasi periodic variations (see e.g. \citealt{anticorrspots}). Gaia EDR3 2101440537737091200 (in the background of KIC 5611382) is a candidate for this type of variation (middle panel of Fig.~\ref{fig:etv}) with a period of $P_{spot}\approx127$ days.

The ETV of binaries may also be caused by a change in the mass ratio, either by mass transfer between the components or mass ejection from the system. A candidate for this is Gaia EDR3 2079688624328740864 (in the background of KIC 8509015) which has an $O-C$ shape that is well-described by a parabolic function (see bottom panel of Fig.~\ref{fig:etv}). If the assumption of the presence of an on-going mass transfer is right, then the orbital period of the system is decreasing with a rate of $\sim0.57$ seconds per year. However, it cannot be ruled out that the $O-C$ diagram represents only a small portion of a long periodic variation caused by a third body.

The full ETV analysis of all of the SPEBs presented in this work will be the subject of a separate paper.

\section{Summary} \label{sec:summary}

We have conducted a survey searching for short period (P $\le$ 12 hours) eclipsing binaries in the background pixels of the observed KIC targets in the original $Kepler$ field. Our main result is that we have found 547 previously unknown, faint SPEBs, mostly W UMa types but also some Algol, $\beta$ Lyrae and ellipsoidal types. This represents a substantial addition to the 605 short period binaries ($0.18 < P < 0.5$ days) already discovered in the \textit{Kepler} field of view and logged in the \textit{Kepler} EB catalogue.

We prepared the light curves for each Quarter and each target for further scientific analysis, that is, we detrended and normalized the Quarters and then combined them to form $\sim$3.5 years long data sets (excluding Quarters 0, 1 and 17). We cross matched the pixel coordinates with the Gaia catalogue to identify the source candidates. In 442 cases we were able to locate a single source for the binary, but for the other 105 there are multiple possible sources. In another 9 cases no candidate source was located close to the pixel coordinates and these were discarded as false positives. We have found that the orbital period distribution of the new binaries is consistent with other authors who have found that the number peaks at around 0.30 days with a sharp cutoff at around 0.20 days.
We have also found that the occurrence rate of short period eclipsing binaries rises with increasing galactic latitude, an unexpected result in the light of the declining occurrence rate reported by \citet{kirk2016}.

In addition, we have used the relationships derived by \citet{Gazeas_Stepien2008} and \citet{Eggleton1983} to estimate some of the physical and geometrical properties of the W UMa type nt-EBs. The obtained parameters can be used as an input for a detailed binary light curve modeling.

These new findings allow us to extend the number of known binaries that show active mass transfer, spots or, most interestingly, have a tertiary component. We constructed the ETV curves and have highlighted some of these. The full results will be presented and analysed in detail in a separate paper. The results for RR Lyrae type pulsating variables and long period eclipsing binaries will also be presented in subsequent papers. It is clear that a wealth of material still remains to be exploited in the \textit{Kepler} observations.

The full set of Q2--Q16 light curves discussed in this paper may be obtained from: \url{https://konkoly.hu/KIK/data_en.html}.

\section{Acknowledgements} \label{sec:ack}

This paper includes data collected by the Kepler mission. Funding for the Kepler mission is provided by the NASA Science Mission directorate.

This project has been supported by the Lend\"ulet Program, project No.  LP2018-7/2020 and the MW-Gaia COST Action (CA18104).

This work has made use of data from the European Space Agency (ESA) mission {\it Gaia} (\url{https://www.cosmos.esa.int/gaia}), processed by the {\it Gaia} Data Processing and Analysis Consortium (DPAC,
\url{https://www.cosmos.esa.int/web/gaia/dpac/consortium}). Funding for the DPAC has been provided by national institutions, in particular the institutions participating in the {\it Gaia} Multilateral Agreement.

We have also made use of the VizieR catalogue access tool, CDS, Strasbourg, France (DOI: 10.26093/cds/vizier). The original description of the VizieR service was published in A\&AS 143, 23.

The research was also carried out with the use of a number of other facilities which we gratefully acknowledge:
\begin{enumerate}
    \item Lightkurve, a Python package for Kepler and TESS data analysis \citep{2018ascl.soft12013L}.
    \item The "Aladin sky atlas" developed at CDS, Strasbourg Observatory, France \citep{2000A&AS..143...33B}.
    \item Astropy,\footnote{http://www.astropy.org} a community-developed core Python package for Astronomy \citep{astropy:2013, astropy}.
\end{enumerate}

\clearpage
\bibliography{JABpaper.bib}{}

\newpage
\appendix
\section{Table of Results}\label{sec:results2}

This section contains the full set of results for the 547 SPEBs identified, including:
\begin{enumerate}
    \item KIC ID: this is the \textit{Kepler} target aperture in which the candidate binary was found. It does not refer to the candidate itself.
    \item Visual Class: the classification given to the binary by visual inspection, i.e. EA: Algol type; EB: $\beta$ Lyrae type; EW: W UMa type; ELL: ellipsoidal type.
    \item Morph. Class: the classification, in the range of 0--1, given to the binary based on the light curve morphology described in section \ref{sub:class}, where high values indicate an ellipsoidal type and low values an Algol type.
    \item $BJD_0$: The zero epoch given in JD-2454833.    
    \item PDM Period: the orbital period obtained using Phase Dispersion Minimisation analysis.
    \item LS Period: the orbital period obtained using Lomb-Scargle analysis.
    \item Gaia Candidate(s): The one or more Gaia candidate sources found for each binary. Note that three of the candidates (Gaia EDR3 2052259348270963328, Gaia EDR3 2077826292150126592 and Gaia EDR3 2078070241990316160 in the backgrounds of KIC 3241227, KIC 7368456 and KIC 7693206 respectively) are marked as variable in the \textit{Gaia} DR2 catalogue.
\end{enumerate}

\pgfplotstableset{
begin table=

\pgfplotstabletypeset[
      multicolumn names, 
      col sep=comma, 
      display columns/0/.style={
		column name=$KIC\ ID$, 
		column type={c},string type},
      display columns/1/.style={
		column name=$Visual$,
		column type={c},string type},
	display columns/2/.style={
		column name=$Morph.$,
		column type={r},string type},
	display columns/3/.style={
		column name=$BJD_0$,
		column type={c},string type},		
	display columns/4/.style={
		column name=$PDM\ Period$,
		column type={r},string type},
	display columns/5/.style={
		column name=$LS\ Period$,
		column type={r},string type},
	display columns/6/.style={
		column name=$Gaia\ Candidates(s)$,
		column type={l},string type},
every head row/.style={before row=\caption{Eclipsing Binary class, period and source data}\label{tab:results}\\\toprule, after row=\\\toprule\endhead}, 
every last row/.style={after row=}]{EDR3_upload_list.csv}


\newpage
\section{Physical and Geometrical Parameters}\label{sec:phys_param}

This section contains the estimated physical and geometric parameters for the 446 W UMa type nt-EBs derived using the period relationships derived by \citet{Gazeas_Stepien2008} and \citet{Eggleton1983}. The table includes:
\begin{enumerate}
    \item KIC ID: this is the \textit{Kepler} target aperture in in the which the candidate binary was found. It does not refer to the candidate itself.
    \item PDM Period: the orbital period obtained using Phase Dispersion Minimisation analysis.
    \item LS Period: the orbital period obtained using Lomb-Scargle analysis.
    \item The estimated absolute $M_V$ magnitude.
    \item The estimated primary and secondary component masses, $M_1$ and $M_2$ (in $M_{\odot}$)
    \item The estimated orbital semi-major axis, $a$ (in $R_{\odot}$)
    \item The mass ratio, $q$, i.e. $M_2/M_1$
    \item the estimated primary and secondary Roche lobe radii, $R_1$ and $R_2$ (in $R_{\odot}$)
\end{enumerate}


\pgfplotstabletypeset[
      multicolumn names, 
      col sep=comma, 
      display columns/0/.style={
		column name=$KIC\ ID$, 
		column type={c},string type},
	display columns/1/.style={
		column name=$PDM\ Period$,
		column type={r},string type},
	display columns/2/.style={
		column name=$LS\ Period$,
		column type={r},string type},
	display columns/3/.style={
		column name=$M_V$,
		column type={r},string type},		
	display columns/4/.style={
		column name=$M_1$,
		column type={r},string type},
	display columns/5/.style={
		column name=$M_2$,
		column type={r},string type},
	display columns/6/.style={
		column name=$a$,
		column type={r},string type},
	display columns/7/.style={
		column name=$q$,
		column type={r},string type},
	display columns/8/.style={
		column name=$R_1$,
		column type={r},string type},
	display columns/9/.style={
		column name=$R_2$,
		column type={r},string type},		
every head row/.style={before row=\caption{Eclipsing Binary Physical Parameters}\label{tab:phys_param}\\\toprule, after row=$ $ & $(days)$ & $(days)$ & $ $ & $(M_{\odot})$ & $(M_{\odot})$ & $(R_{\odot})$ & $ $ & $(R_{\odot})$ & $(R_{\odot})$\\\toprule\endhead}, 
every last row/.style={after row=}]{M_R_table.csv}


\end{document}